\documentclass[10pt,preprint]{aastex}
\slugcomment{Not to appear in Nonlearned J., 45.}

\shorttitle{Solar Radio Emission}
 \shortauthors{Tan et al.}

\begin{document}


\title{Study of Calibration of Solar Radio Spectrometers and the quiet-Sun Radio Emission}


\author{Chengming Tan\altaffilmark{1}, Yihua Yan\altaffilmark{1}, Baolin Tan\altaffilmark{1}, Qijun Fu\altaffilmark{1}, Yuying Liu\altaffilmark{1}}
\affil{\altaffilmark{1}Key Laboratory of Solar Activity, National
Astronomical Observatories of Chinese Academy of Sciences, Datun
Road A20, Chaoyang District, Beijing 100012, China}

\author{Guirong Xu\altaffilmark{2}}
\affil{\altaffilmark{2} Hubei Key Laboratory for Heavy Rain
Monitoring and Warning Research, Institute of Heavy Rain, China
Meteorological Administration, Wuhan 430205, China}

\begin{abstract}
This work presents a systematic investigation of the influence of
weather conditions on the calibration errors by using Gaussian
fitness, least chi-square linear fitness and wavelet transform to
analyze the calibration coefficients from observations of the
Chinese Solar Broadband Radio Spectrometers (at frequency bands of
1.0-2.0 GHz, 2.6-3.8 GHz, and 5.2-7.6 GHz) during 1997-2007. We
found that calibration coefficients are influenced by the local
air temperature. Considering the temperature correction, the
calibration error will reduce by about $10\%-20\%$ at 2800 MHz.
Based on the above investigation and the calibration corrections,
we further study the radio emission of the quiet-Sun by using an
appropriate hybrid model of the quiet-Sun atmosphere. The results
indicate that the numerical flux of the hybrid model is much
closer to the observation flux than that of other ones.
\end{abstract}


\section{Introduction}

Broadband radio spectrometers play an important role in observing
and revealing the physical processes of the solar atmosphere,
solar flares, coronal mass ejections, and other solar activities.
When we utilize the data observed by spectrometers to study solar
problems, calibration is a key procedure that dominates the
reliability of the observational results. At present, there are
several solar broadband radio spectrometers running in the world,
such as Phoenix at ETH Zurich (100-4000 MHz; \citet{Benz91}),
Ondrejov Radiospectrograph in the Czech Republic (800-5000 MHz;
\citet{Jiricka93}), Brazil Broadband Spectrometer (200-2500 MHz;
\citet{Sawant01}) and the Chinese Solar Broadband Radio
Spectrometers (SBRS, 1.10-2.06, 2.60-3.80, and 5.20-7.60 GHz;
\citet{Fu95}, \citet{Ji05}) at the Huairou Solar Observing
Station. All the above spectrometers are single-dish telescopes
that receive the radio emission of the full solar disk without
spatial resolutions. So far, calibration techniques of single-dish
radio telescopes can be classified as follows:

(1) Absolute calibration. Skillful practices are needed to
determine the absolute calibration values by using an approximate
gain of the antenna based on radar and communication techniques,
together with an approximate sensitivity derived from experiments
with limited accuracy. It is still a difficult problem and usually
only applied to polarimeters (\citet{Broten60}; \citet{Findlay66};
\citet{Tanaka73}; etc).

(2) Relative calibration. This method needs reference emission
sources to calculate the emission flux (or brightness
temperature). It is usually applied to spectrometers since the
absolute calibration of spectrometers would be a huge and very
complex mission at broadband frequencies \citep{Messmer99}.

(3) Nonlinear calibration. When the gain factor of the receiver
exceeds its normal range during strong solar radio burst, it is
necessary to consider nonlinear calibration \citep{Yan02}.

Recently, we found that the calibration data has a strong
correlation with the local air temperature at frequency of 2.6 -
3.8 GHz of SBRS \citep{Tan09}. The local air temperature is an
important factor that needs to be taken into account in the
calibration procedure. We also analyze the impact of the Sun
elevation and other weather conditions on the calibration.

Moreover, the calibration of solar radio spectrometers is very
important when we compare the observations with theoretical work
in the study of the quiet-Sun radio emission. It supplies an
accurate flux spectrum, which help us to study the basic nature of
the quiet-Sun emission and provide a fundamental knowledge of the
solar atmosphere. The radio quiet-Sun has been studied for about
70 yr. \citet{Martyn46} considered the quiet-Sun as a black body
radiation and studied its radio spectrum. He showed the variation
of radio brightness across the solar disk at various frequencies
and indicated that limb brightening should be observed.
\citet{Smerd50} studied the radio radiation from the quiet-Sun
with numerical analysis by applying radiative transfer equations
to a typical ray trajectory \citep{Jaeger50}. He presented a
complete analysis of the quiet-Sun radio radiation, and the result
is consistent with the observations at wavelengths of 3 cm, 10 cm,
25 cm, 50 cm, 60 cm, 1.5 m \& 3.5 m (\citet{Pawsey49};
\citet{Chris51}). Since then, more and more observations and
studies have been published (\citet{Alllen57}; \citet{Tanaka73};
\citet{Bastian96}). \citet{Kundu65} reviewed the basic theory of
radiative transform and propagation in solar radio astrophysics.
\citet{Tanaka73} used the average value of the daily noon flux to
obtain the radio flux spectrum of the quiet-Sun. Other works
(\citet{Nelson85}; \citet{Zirin91}) used the flux density of the
quiet-Sun around the minimum of the solar cycle. The numerical
computation of the quiet-Sun radio emission has been discussed in
those papers (\citet{Dulk85}; \citet{Benz93}; \citet{Selhorst05}).
\citet{Benz09} pointed out that the radio emission of the
quiet-Sun is a well-defined minimum radiation level when the Sun
has no sunspots for some weeks. The presence of sunspots enhances
the radio emission and produces a slowly variable component.
\citet{Shibasaki11} reviewed radio emission of the quiet-Sun in
observations and numerical computation. However, until now few
works have paid attention to the study of the totally quiet-Sun
radio flux spectrum at centimeter to meter wavelengths. Our work
will contribute new observations and computations on this topic by
improving the study in several aspects: (1) adopt new flux
spectrum observations during solar cycle 23; (2) select the daily
noon flux without sunspots $\pm3$ days as the flux of the
quiet-Sun; (3) utilize numerical integration instead of an
approximate analytical function in numerical computation of radio
emission; (4) use three models (\citet{Vernazza81};
\citet{Selhorst05}; \citet{Fontenla11}) of the solar atmosphere
for comparison of numerical computations. We do not consider the
magnetic field $B$ and the scattering effect since the magnetic
field $B$ is weak in the quiet-Sun and the scattering effect can
be neglected at shorter meter wavelengths (\citet{Aubier71};
\citet{Thejappa08}).

SBRS has observed the sun since 1994 and accumulated observation
data for more than one solar cycle. In this work, we take the
observation data of SBRS to analyze the daily calibration and
increase the accuracy of the calibration procedure in section 2.
Based on the radio flux spectrum of the quiet-Sun, section 3
presents numerical analysis of the quiet-Sun radio emission with
improved numerical treatment and several new models of the solar
atmosphere. Section 4 gives the main conclusion and some
discussions.

\section{Observations and Calibration}

Three spectrometers of SBRS (Fu et al. 1995, 2004; \citet{Ji05})
are located at the Huairou Solar Observing Station in China. The
first spectrometer (SBRS1) was upgraded three times. SBRS1
observed the Sun and only saved the burst data in the early years.
It started routine and calibration observations in 1999 October .
It was under construction after 2006 July and started to work
again in 2013. The second spectrometer (SBRS2) began to observe
the Sun in 1996 September. The third spectrometer (SBRS3) began
observing the Sun in 1999 August. SBRS2 and SBRS3 shared the same
antenna, with a diameter of 3.2 meter. The main information and
performed parameters of SBRS are listed in table \ref{tbl-0}.
During solar cycle 23, the SBRS observed the Sun and did daily
calibration at each noon (avoiding the effect of radio bursts),
except for the instruments under maintenance. They supply plenty
of observations and calibration data to analyze the quiet-Sun
radio emission. In this work, we do not consider the daily
calibration data in less than 1 yr of observations because small
data sets may result in larger error. We only select the daily
calibration data from October 1999 to October 2004 of SBRS1, from
1997 to 2007 of SBRS2, and from 2000 to 2007 of SBRS3 except for
some big data gaps ($>10days$) or without a noise source or
termination (Left panel of Fig.\ref{figsystem}). We go on analyze
the daily calibration data with the daily noon flux from the
National Geophysical Data Center (NGDC), and we study some
treatments to reduce the calibration errors. At last, the flux
spectrum of the quiet-Sun can be obtained from the observing data
with calibration.

\begin{table}[ht]
\begin{center}
\caption{The main information and performance parameters of the
component spectrometers \label{tbl-0}}
\begin{tabular}{crrrrrrr}
\tableline\tableline
   Band     &  Dm        &  $f_{re}$       & Cadence    & Pol         & Chan      & Observation Period \\
  (GHz)     & (m)        & (MHz)       & (ms)       &             &           &     \\
\tableline
%  Freqency & Diameter   & Frequency  & Temporal   & Polarization& Number of & Observation\\
%  range    & of antenna & resolution & resolution & accuracy    & channel   & period\\
 SBRS1 (1.0-2.0)  & 7.3        & 20.0       & 100        & $<$10$\%$    & 50        & 1994-2002 Jan \\
 SBRS1 (1.10-2.06)& 7.3        & 4.0        & 5          & $<$10$\%$    & 240       & 2002 May-2004 Oct 25 \\
 SBRS1 (1.10-1.34)& 7.3        & 4.0        & 1.25       & $<$10$\%$    & 240       & 2004 Oct 26-2006 Jul \\
 SBRS1 (1.10-2.10)& 7.4        & $\sim$2.78    & 5       & $<$10$\%$    & 360       & 2013 Aug 20-now \\
 SBRS2 (2.6-3.8)  & 3.2        & 10.0       & 8 or 2$^{*}$  & $<$10$\%$    & 120       & 1996 Sep-now \\
 SBRS3 (5.2-7.6)  & 3.2        & 20.0       &5 or 1.25$^{*}$& $<$10$\%$    & 120       & 1999 Aug-now\\
\tableline
\end{tabular}
\tablecomments{The first column 'Band' is the frequency band of
the spectrometer. 'Dm' is the diameter of the antenna. '$f_{re}$'
is the frequency resolution. 'Cadence' is the time cadence. 'Pol'
is the circular polarization accuracy. 'Chan' is channel number.}
 \tablenotetext{*}{the temporal resolution when the spectrometer works at a quarter of frequency band.}
\end{center}
\end{table}

\subsection{Fundamental of Calibration}

\begin{figure*}[ht]
\epsscale{1.0} \plotone{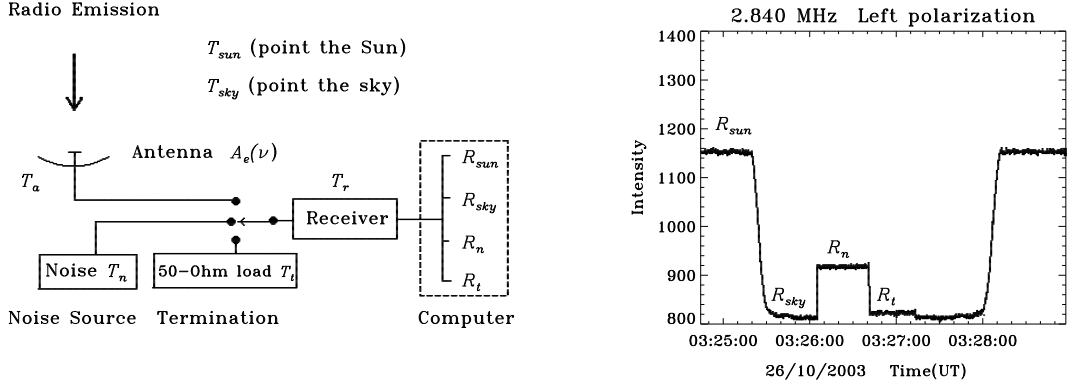}
 \caption{Left panel is the
schematic chart of the antenna and receiver system. Right panel is
a figuration of the daily calibration data. \label{figsystem}}
\end{figure*}

Since our data are obtained by broadband spectrometers, this work
mainly focus on the relative calibration. The basic principle of
the relative calibration was discussed in \citet{Messmer99} and
\citet{Yan02} in detail. SBRS records a set of calibration data in
daily routine observations. The left panel of Fig.\ref{figsystem}
presents the schematic chart of the antenna and receiver system.
The switch may turn to the antenna, noise source, and termination,
respectively. Usually, the termination is a 50 ohm resistor. For
each case there is an input to the receiver system. $T_{a}$ is
equivalent antenna temperature ($T_{a}=T_{sun}$ or $T_{a}=T_{sky}$
when the antenna points to the Sun or the sky, respectively).
$T_{n}$ and $T_{t}$ are the equivalent temperature of the noise
source and the terminal source, respectively. Therefore, there are
four outputs at the computer at various frequencies: the Sun
(marked as $R_{sun}$), sky background ($R_{sky}$), noise source
($R_{n}$), and terminal source ($R_{t}$). They are showed in the
right panel of Fig.\ref{figsystem} by the data processing
software. When the gain factor ($Gr$) of the receiver is in the
linear range \citep{Yan02}, they comply with the following
relationships:

\begin{equation}
R_{sun}(\nu)=(T_{sun}+T_{r}(\nu))\cdot Gr(\nu) \label{calisun}
\end{equation}
\begin{equation}
R_{sky}(\nu)=(T_{sky}+T_{r}(\nu))\cdot Gr(\nu) \label{calisky}
\end{equation}
\begin{equation}
R_{n}(\nu)=(T_{n}(\nu)+T_{r}(\nu))\cdot Gr(\nu) \label{calinoise}
\end{equation}
\begin{equation}
R_{t}(\nu)=(T_{t}(\nu)+T_{r}(\nu))\cdot Gr(\nu) \label{caliterm}
\end{equation}

Here $\nu$ is the observed frequency, and $T_{r}$ is equivalent
temperature of the receiver system, which is very small. From
\citet{Broten60} and \citet{Messmer99}, we have,

\begin{equation}
F_{sun}(\nu)=\frac{2k_{B}\cdot T_{sun}(\nu)}{A_e(\nu)}
\end{equation}

$F_{sun}$ and $F_{sky}$ are the radio fluxes of the quiet-Sun and
sky, respectively. The effective area of the antenna aperture
$A_e(\nu)$ is changeless and can be considered as a constant.
$k_{B}$ is Boltzmann's constant. The real flux of the sun
$F_{0}(\nu)$ should subtract the contribution of the sky
$F_{sky}(\nu)$, i.e., equation (\ref{puresun}). From equation
(\ref{calisun}) and (\ref{calisky}) we have:

\begin{equation}
F_{0}(\nu)=
F_{sun}(\nu)-F_{sky}(\nu)=\frac{2k_{B}}{A_e(\nu)}\cdot\frac{R_{sun}(\nu)-R_{sky}(\nu)}{Gr(\nu)}
\label{puresun}
\end{equation}

$Gr(\nu)$ can be deduced with equation (\ref{calinoise}) and
(\ref{caliterm}). Then equation (\ref{puresun}) can be transformed
into

\begin{equation}
F_{0}(\nu)=\frac{2k_{B}\cdot(T_{n}(\nu)-T_{t}(\nu))}{A_{e}(\nu)}\cdot
\frac{R_{sun}(\nu)-R_{sky}(\nu)}{R_{n}(\nu)-R_{t}(\nu)}
 \label{caliobs}
\end{equation}

where $R_{sun}(\nu)$, $R_{sky}(\nu)$, $R_{n}(\nu)$ and
$R_{t}(\nu)$ are daily calibration data. As written previously,
$k_{B}$ and $A_e(\nu)$ are constant. $T_{n}$ and $T_{t}$ are
equivalent temperatures of the noise source and the terminal
source, respectively. The noise source and the terminal source are
fixed electronic apparatus and usually stable under the steady
environment and no interference. Thus, in equation (\ref{coeff}),
$C(\nu)$ will be also stable under the steady environment and no
interference. It is defined as calibration coefficient

\begin{equation}
C(\nu)=2k_{B}\cdot\frac{(T_{n}(\nu)-T_{t}(\nu))}{A_{e}(\nu)}=F_{0}(\nu)\cdot\frac{R_{n}(\nu)-R_{t}(\nu)}{R_{sun}(\nu)-R_{sky}(\nu)}
\label{coeff}
\end{equation}

For any observed data $R_{sun}(\nu)$ , the corresponding real flux
$F_{\odot}(\nu)$ of the Sun is calibrated as follows:

\begin{equation}
F_{\odot}(\nu)=
F_{sun}(\nu)-F_{sky}(\nu)=\frac{R_{sun}(\nu)-R_{sky}(\nu)}{R_{n}(\nu)-R_{t}(\nu)}\cdot
C(\nu) \label{calisfu}
\end{equation}

Equation (\ref{calisfu}) is very convenient to do calibration for
it does not require that we know the values of
$T_{n}(\nu)-T_{t}(\nu)$ and $A_{e}(\nu)$. The bandpass flatness of
the spectrum is usually not good for two reasons: 1) the gain
factor $Gr(\nu)$ of the receiver varies along frequency; and 2)
the antenna system $A_{e}(\nu)$ will also vary more or less along
frequency. The calibration with equation (\ref{calisfu}) will
eliminate the frequency property of the gain factor and flat the
frequency property of the antenna system with $C(\nu)$ of each
frequency. The most important of calibration work is to analyze a
suitable calibration coefficient $C(\nu)$ with the right-hand side
of expression (\ref{coeff}) and reduce the calibration error.
Usually we decide the daily noon flux from NGDC as $F_{0}(\nu)$.
It can be downloaded from the NGDC Web
site\footnote{http://www.ngdc.noaa.gov/stp/SOLAR/ftpsolarradio.html},
which provides standard flux of solar radio emission at nine fixed
frequencies (245, 410, 610, 1415, 2695, 2800, 4995, 8800,
15400MHz). We can calculate flux at any frequency between 245 and
15,400 MHz in using of linear interpolation.

\subsection{Impact of the weather in the calibration}

\begin{figure*}[ht]
\epsscale{1.00} \plotone{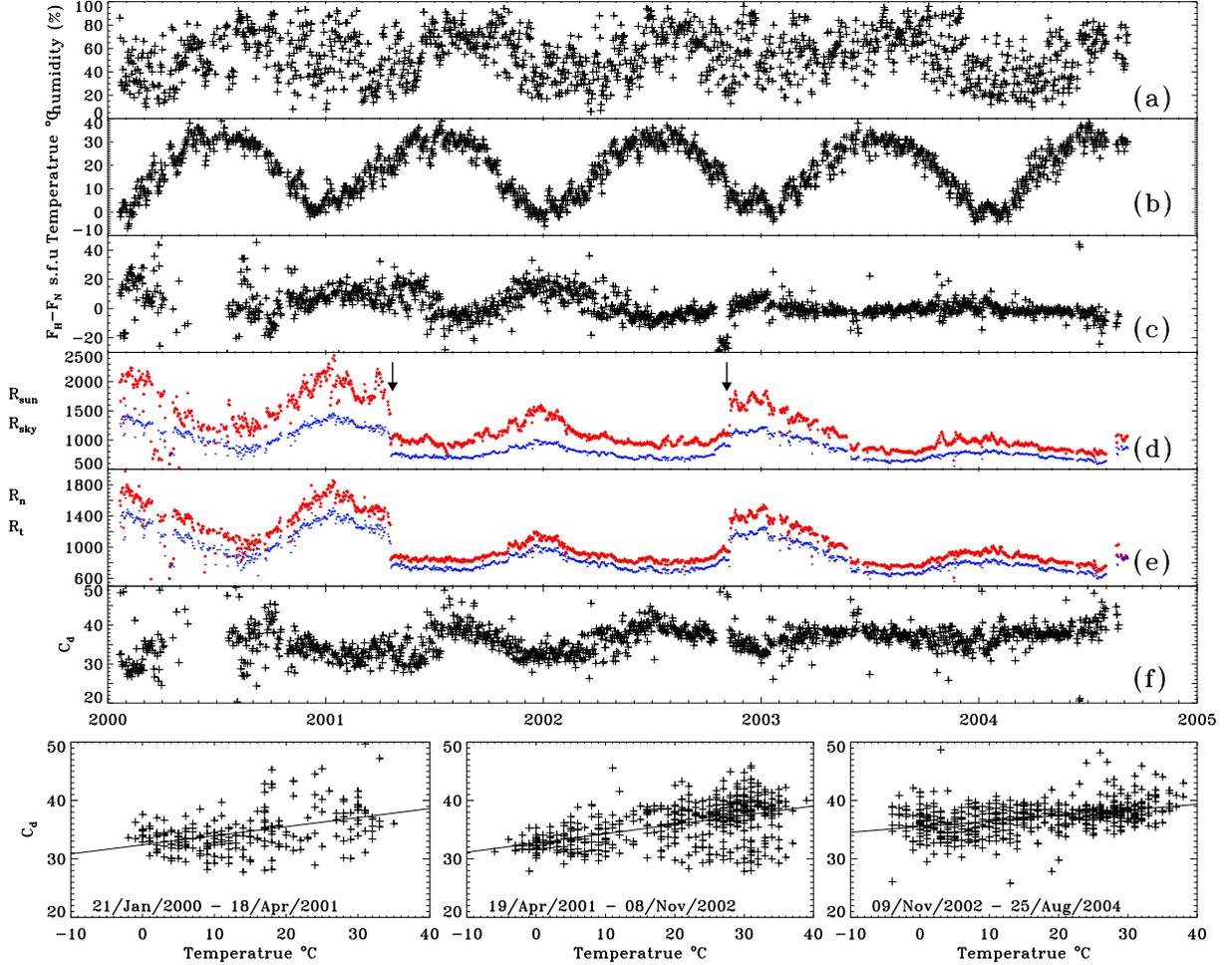}
 \caption{(a and b) Local humidity and local
air temperature, respectively; (c) difference between the daily
noon flux of SBRS (calibrated with $\overline{C}$) and the daily
noon flux from NGDC; (d) $R_{sun}$ (red small cross) and $R_{sky}$
(blue dotted); (e) $R_{n}$ (red small cross) and $R_{t}$ (blue
dotted); (f) daily calibration coefficients
$C_{d}=F_{0}\cdot\frac{R_{n}-R_{t}}{R_{sun}-R_{sky}}$. The bottom
three panels plot the daily calibration coefficients vs. local air
temperature for three terms. The sloping line is the least
chi-square linear fitness of $C_{d}$ and the local air
temperature. All the radio data are at 2800 MHz. \label{figcali1}}
\end{figure*}

Generally, a constant coefficient is adopted in calibration. In
practice, the calibration is influenced by the local air
temperature more or less. Panels (b) and (c) of Fig.\ref{figcali1}
show the comparison between the local air temperature and the
calibration result at 2800 MHz. Panels (d) and (e) of
Fig.\ref{figcali1} show that $R_{sun}(\nu)$, $R_{sky}(\nu)$,
$R_{n}(\nu)$, and $R_{t}(\nu)$ are all correlated with the local
air temperature. From equation (\ref{calisun}) and (\ref{calisky})
we can deduce that the gain factor $Gr(\nu)$ is influenced by the
local air temperature since $T_{sun}(\nu)$ and $T_{sky}(\nu)$ have
no relationship with air temperature and $T_{r}(\nu)$ is very
small. In fact, the gain factor $Gr(\nu)$ of all bands of SBRS is
influenced by the local air temperature more or less. Panels (d)
and (e) also show two major jumps (arrows in the figure) at 2001
April 19 and 2002 November 09, respectively. We check the daybook
of the observer and find that there is a change to the attenuation
of the instrument. Panel (f) shows that the daily calibration
coefficients $C_{d}$ are correlated with the local air
temperature. The bottom three panels of Fig.\ref{figcali1}
indicate that the correlation between $C_{d}$ and the local air
temperature varied because adjustment of the instrument. We can
deduce that ${T_{n}(\nu)-T_{t}(\nu)}$ in equation (\ref{coeff}) is
also influenced by the local air temperature. All these results
imply that the electronic apparatuses of the instruments are
influenced by the local air temperature. The relationship between
the calibration result (panel (c) of Fig.\ref{figcali1}) and the
humidity (panel (a) of Fig.\ref{figcali1}) or other weather
conditions is not clear. The observation must be stopped when some
extreme weather events occur (such as thunder, windstorms,
rainstorms and heavy snow). Until now we have found no distinct
evidence that the calibration was influenced by normal weather
conditions except for air temperature.

We will explain that the Sun elevation has a small effect on the
calibration of this work. \citet{Tsuchiya65} pointed out that
solar flux can be measured without considering the weather
condition (atmospheric absorption) at frequency lower than 17 GHz.
Considering the atmospheric absorption, the antenna temperature is
given by equation as follow.

\begin{equation}
T_{a}=T_{source}e^{-\tau_{0}}+T_{sky} \label{tempabsorp}
\end{equation}
\begin{equation}
T_{sky}=\int_{0}^{\tau_{0}}T(\tau)e^{-\tau}d{\tau}
\label{tempabsorp}
\end{equation}

$T_{source}$ is the real temperature at the emission source.
$T(\tau)$ is the temperature along the line of sight. $\tau$ is
the optical depth of the atmospheric absorption.

\begin{equation}
\tau=\int_{h}^{\infty}{\kappa}{\cdot}dh{\cdot}{sec(z)}
\label{optdepth}
\end{equation}

where $\kappa$ is the absorption coefficient of the atmosphere. It
is small ($<0.006/km$) at frequencies lower than 17 GHz
\citep{Tsuchiya65}. Moreover, the density of the water vapor and
oxygen decreases rapidly along the height of atmosphere $<10km$
above the ground level. The absorption $\kappa$ at the height of
$\sim10km$ will be about $1\%$ of that of the ground level. Here
$h$ is height and $z$ is the zenith distance. So the Sun elevation
is equal to $90^{o}-z$. The daily calibration observation is done
around the noon. The Sun elevation at noon is larger than
$26^{o}$, considering that the latitude of Beijing is $40^{o}$.
The $sec(z)$ is smaller than 5.8 when the Sun elevation is greater
than $10^{o}$. Thus the optical depth in equation (\ref{optdepth})
is also very small, $\tau_{0}\leq0.02$. It has only a small effect
on the temperature of the emission source.

We also find no distinct evidence that the calibration was
influenced by the Sun elevation. If the Sun elevation has a
considerable effect on the observation, the following will result:
(1) the variation profile of $R_{sun}$(or $R_{sky}$) will show
considerable difference from that of $R_{n}$ (or $R_{t}$). The
left panel of Fig.\ref{figsystem} shows that $R_{n}$ and $R_{t}$
have no relationship with Sun elevation for no connection with
antenna; (2) $R_{sun}$(or $R_{sky}$) in summer (high elevation)
will be larger than that in winter (low elevation). This is
opposite to the observation shown in panel (d) of
Fig.\ref{figcali1}. From all the above analyses, the Sun elevation
does not need to be considered in calibration.

We need to understand two problems: (1) the variety of daily
calibration data and (2) the relationship between the daily
calibration coefficients and the local air temperature. The first
is related to the daily calibration data and daily standard flux
of solar radio emission. The daily calibration data was recorded
by SBRS. They consist of 4 recorded parameters: $R_{sun}$,
$R_{sky}$, $R_{n}$, and $R_{t}$ (right panel of
Fig.\ref{figsystem}). The daily noon flux of radio emission at
nine fixed frequencies can be downloaded from the NGDC Web site.
Usually the observations from Learmonth Observatory are considered
since its recording time (05:00 UTC) is very closed to that of
SBRS (04:00 UTC). The second problem is related to the local air
temperature data of Beijing, which are collected from China
Meteorological Administration.

The daily calibration coefficients are calculated by the
right-hand side of equation (\ref{coeff}) with the daily
calibration data of SBRS during 1997 - 2007. There are several
steps of pretreatment to exclude the abnormal data: (1) Look up
the notebook of the observer, check out the day when the
instrument was not in good operation (testing, maintenance, no
noise source, etc.), and rule out the data during these cases; (2)
get rid of the calibration data with interference, $\pm$NaN value
or minus value. The data are wrong and will result in
unpredictable errors in the mathematical processing. The main
reason for the $\pm$ NaN, minus, and big value of the daily
calibration coefficient comes from the electromagnetism
interference of the environment. So $R_{sky}$ will be too big or
even larger than $R_{sun}$. The relationship between the abnormal
data and bad weather conditions is still not clear. We will list
all the possible reasons in section 4.

\begin{figure*}[ht]
\epsscale{1.00} \plotone{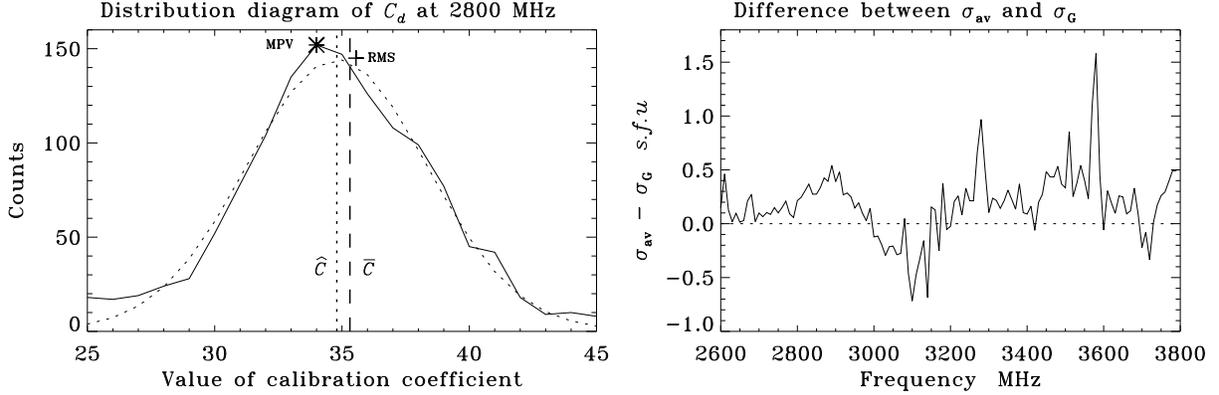}
 \caption{Left: comparison between
the average value $\overline{C}$ (long-dashed line) and the
gaussian fitness value $\widehat{C}$ (dashed line). Right:
difference between the standard deviation of calibration of
$\overline{C}$ and that of $\widehat{C}$.\label{figdiff}}
\end{figure*}

\subsection{Analysis of the Calibration}

We take an example to analyze the calibration at 2800 MHz. At
first, we compare the average value of the daily calibration
coefficients ($\overline{C}$) with the Gaussian fitness value of
the daily calibration coefficients ($\widehat{C}$) at each
frequency. They are $C(\nu)$ in equation (\ref{calisfu}) when
doing calibration. The Gaussian fitness value can exclude the
contribution of big or very small values, which usually indicate
radio bursts, sunspots or abnormal observations. The left panel of
Fig.\ref{figdiff} shows the comparison between $\overline{C}$ and
$\widehat{C}$ at 2800 MHz. The vertical long-dashed line indicates
values of $\overline{C}$. The black solid line is the histogram of
the daily calibration coefficient. The dashed line is the Gaussian
fitness of the black solid line. The Gaussian fitness is:

\begin{equation}
f(x)=A_{0}e^{\frac{-z^{2}}{2}}+A_{3}+A_{4}x+A_{5}x^{2},
z=\frac{x-A_{1}}{A_{2}}
\end{equation}

The $x$ value of the maximum $f(x)$ is $\widehat{C}$ marked as
vertical dashed line in Fig.\ref{figdiff}. When
$|\overline{C}$-$\widehat{C}| \leq$ 0.02*$\overline{C}$, the
average value is close to the gaussian fitness value. This
indicates that the observation of this frequency is reliable. In
fact, $\overline{C}$ and $\widehat{C}$ are close for most of the
frequencies. The most probable value (MPV) is the maximum of the
histogram (marked as star in Fig.\ref{figdiff}). The rms value is
the square root of $C_{d}$ (marked as cross in Fig.\ref{figdiff}).
Both the calibration error of the MPV and the rms are lager than
those of $\overline{C}$ and $\widehat{C}$. The right panel of
Fig.\ref{figdiff} shows the comparison between the standard
deviation ($\sigma_{av}$) of calibration with $\overline{C}$ and
the standard deviation ($\sigma_{G}$) of calibration with
$\widehat{C}$. The standard deviation of calibration is calculated
by equation (\ref{xigma}).

\begin{equation}
\sigma=stddev(F_{H}-F_{N}) \label{xigma}
\end{equation}

Here $F_{H}$ is the daily noon flux calculated by equation
(\ref{calisfu}) with the daily calibration data of SBRS and
different $C$ ($\overline{C}$, $\widehat{C}$, etc.). $F_{N}$ is
the daily noon flux from NGDC. For $82\%$ of frequencies,
$\sigma_{av}$ are greater than $\sigma_{G}$. For the remaining
$18\%$ of frequencies, $\sigma_{av}$ are less than or equal to
$\sigma_{G}$ because the gaussian fitness deviates from the
histogram of the daily calibration coefficients. Thus,
$\widehat{C}$ is used as the constant coefficient of calibration.
The calibration with a constant coefficient ($\widehat{C}$ or
$\overline{C}$) is named the constant calibration.

\begin{figure*}[ht]
\epsscale{1.0} \plotone{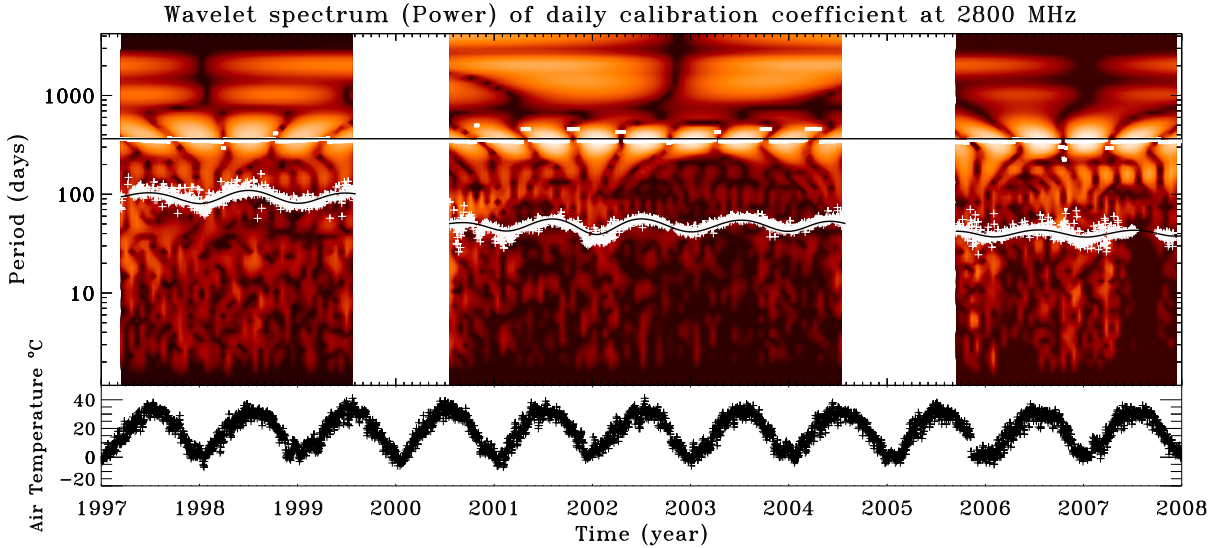} \caption{Top: wavelet spectrum
of daily calibration coefficients at 2800 MHz. The x-axis is the
time (year), and y-axis is the period (days). The white cross is
the daily calibration coefficients. The black smooth curve is the
calibration coefficients after wavelet filtering. The black solid
line shows the period of 365 days. The short white line near the
black line of 365 days is the location of maximum power. The data
gap represents no observation or during maintenance. Bottom: local
air temperature.\label{figwave}}
\end{figure*}

The other two sets of calibration coefficients are related to the
local air temperature, including calibration coefficients with
temperature correction (TC) and temperature-wavelet correction
(TWC). Fig.\ref{figwave} shows a strong correlation between the
daily calibration coefficients (white cross in top panel) at 2800
MHz and local air temperature (black cross in bottom panel) during
1997 - 2007. Here we partition the data into three terms (1997 -
1999, 2000 - 2004, 2005 - 2007) according to the data gap during
which the spectrometers are under maintenance. Those points that
exceed the normal range during maintenance should be excluded.
Then we decide in equation (\ref{f10}) the linear relationship
between the daily calibration coefficients $C_{d}(t)$ and the
local air temperature $T_{air}(t)$ for each term. The variable $t$
is the time count in days. $C_{1}$ and $C_{2}$ can be obtained by
least chi-square linear fitness between $C_{d}(t)$ and
$T_{air}(t)$:

\begin{equation}
C_{d}(t)=C_{1}+C_{2}*T_{air}(t) \label{f10}
\end{equation}

Here $C_{1}$ is the first term of the linear fitness. In the
second term of the linear fitness, $C_{2}$ is the correct factor
that is related to the local air temperature. For each day, the
calibration coefficient $C_{TC}=C_{1}+C_{2}*T_{air}$ can be
corrected by the local air temperature and replace $C(\nu)$ in
equation (\ref{calisfu}). This is the calibration with temperature
correction TC.

TWC is the calibration coefficient corrected by
temperature-wavelet analysis. The top panel of Fig.\ref{figwave}
plots the power spectrum (red color) of $C_{d}(t)$ after wavelet
transform for three terms. The wavelet transform method was
discussed by \citet{Sy02} in detail. Their work used a complex
basis based on the Morlets wavelet which is well localized in both
the scale and frequency plane. This makes it possible to rapidly
reconstruct the signal even in the presence of a singularity value
or in the absence of data. We first transform $C_{d}(t)$ by
wavelet to the wavelet data, which are complex numbers and give
the power information along the time and period. The period of the
maximum power is about 365 days, and the half-power beam widths
(HPBWs) are in the range of 260$\sim$530 days. We go on to set the
value of the wavelet data beyond the HPBW as zero (that is, we
only keep the wavelet data within the HPBWs), and we perform
wavelet transform back to new coefficients $\widetilde{C}(t)$.
This numerical process is wavelet filtering, which filters out the
information we do are not concerned with. The new coefficients
$\widetilde{C}(t)$ are plotted as a black smooth curve through
$C_{d}(t)$ (white cross). In equation (\ref{linfittwc}),
$\widetilde{T}_{air}(t)$ is the local air temperature after
wavelet filtering. So, $C_{1}$ and $C_{2}$ can be obtained by
least chi-square linear fitness between $\widetilde{C}(t)$ and
$\widetilde{T}_{air}(t)$:

\begin{equation}
\widetilde{C}(t)=C_{1}+C_{2}*\widetilde{T}_{air}(t)
\label{linfittwc}
\end{equation}

$C_{TWC}=C_{1}+C_{2}*T_{air}$ can replace $C(\nu)$ in equation
(\ref{calisfu}) when doing calibration. This is the calibration
with TWC.

\begin{figure*}[ht]
\epsscale{1.00} \plotone{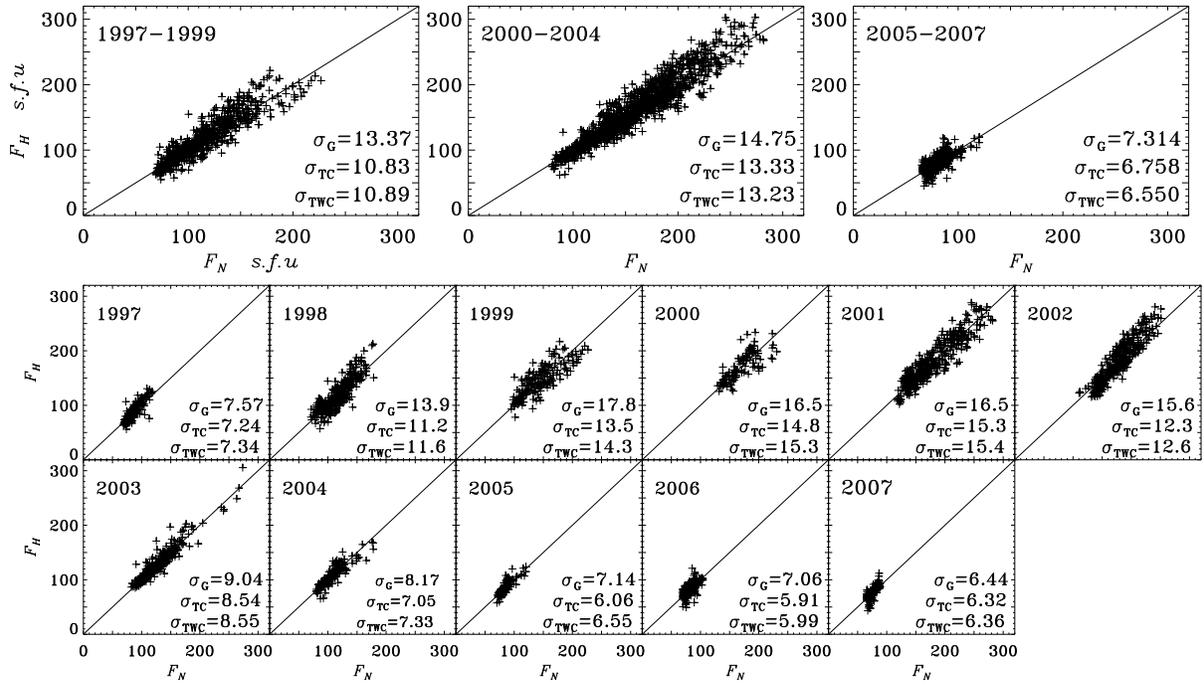} \caption{Each panel plots the
daily noon flux of SBRS after $\widehat{C}$ calibration vs. daily
noon flux of NGDC. They are plotted as black crosses. The top
three panels are the comparison of calibration results with
$\widehat{C}$, $C_{TC}$ and $C_{TWC}$ at 2800 MHz. The bottom 11
panels are the yearly calibration results with $\widehat{C}$,
$C_{TC}$ and $C_{TWC}$ at 2800 MHz.\label{stdcompare}}
\end{figure*}

We compare the calibration results of three sets of calibration
coefficients ($\widehat{C}$, $C_{TC}$ and $C_{TWC}$) at 2800 MHz
in both the long term (more than 2 yr) and short term (1 yr). The
standard deviations  $\sigma_{G}$, $\sigma_{TC}$, and
$\sigma_{TWC}$ are of calibration with $\widehat{C}$, $C_{TC}$ and
$C_{TWC}$, respectively. They are calculated by equation
(\ref{xigma}). Each panel of Fig.\ref{stdcompare} plots the daily
noon flux observed by SBRS (calibration with $\widehat{C}$) versus
the flux of NGDC. The upper three panels are the calibration
results for three long observation terms. The bottom 11 panels
show the yearly calibration results. The $\sigma_{TC}$ or
$\sigma_{TWC}$ are about $10\%\sim20\%$ smaller than $\sigma_{G}$
at various observation terms. The difference between $\sigma_{TC}$
and $\sigma_{TWC}$ is small, $\sim2\%$. In short observation
terms, $\sigma_{TC}$ is the smallest. While in long observation
terms, $\sigma_{TWC}$ is a little smaller than $\sigma_{TC}$ in
most cases. We conclude that TC calibration is better for short
observation terms, while TWC calibration is better for long
observation terms. The relative standard deviations (RSDs) of
calibration with $C_{TC}$ and $C_{TWC}$ are less than $10\%$ in
all cases, while RSDs of calibration with $\widehat{C}$ are
greater than $10\%$ sometimes because it cannot correct the
influence of the air temperature. The RSD of the calibration will
be larger when the influence of air temperature is stronger. The
$C_{TC}$ of 2007 is used in the calibration during the years of
2008 and 2009. At 2800MHz, $\sigma_{TC}=6.9 s.f.u$ is about the
same as for the years of 2004 - 2007. In practice, the calibration
coefficient $C(\nu)$ should be updated annually.

The daily calibration data of SBRS1 and SBRS3 also have a strong
relationship with the local air temperature. The relationship
between $C_{d}$ and the local air temperature is quite complex
because SBRS1 and SBRS3 were examined and repaired many times. In
short observing terms less than 1 yr, there is no significant
difference ($<5\%$) between constant calibration and TC (or TWC)
calibration. The constant coefficients are used in calibration of
them. At the 1.0-2.0 GHz band, the daily noon flux of SBRS1 at
1420 MHz is compared with the daily noon flux of NGDC data at 1415
MHz. The standard deviations are varied within $9\sim12 s.f.u$ in
different observation terms. The same is done at 1.10-2.06 GHz
band. The standard deviations of calibration are varied within
$8\sim12 s.f.u$ at 1420 MHz in different observation terms. At the
5.2-7.6 GHz band, there are no observation data from NGDC. The
linear interpolated values between 4995 and 8800 MHz of NGDC are
used as comparison. The standard deviations of calibration are
varied within $5\sim18 s.f.u$ at 5900 MHz in different observation
terms. The observation term with the smallest standard deviation
indicates a stable system and few occasional interferences,
therefore signifying the best status of the instrument.

\subsection{Observation Results}

\begin{figure*}[ht]
\epsscale{1.0} \plotone{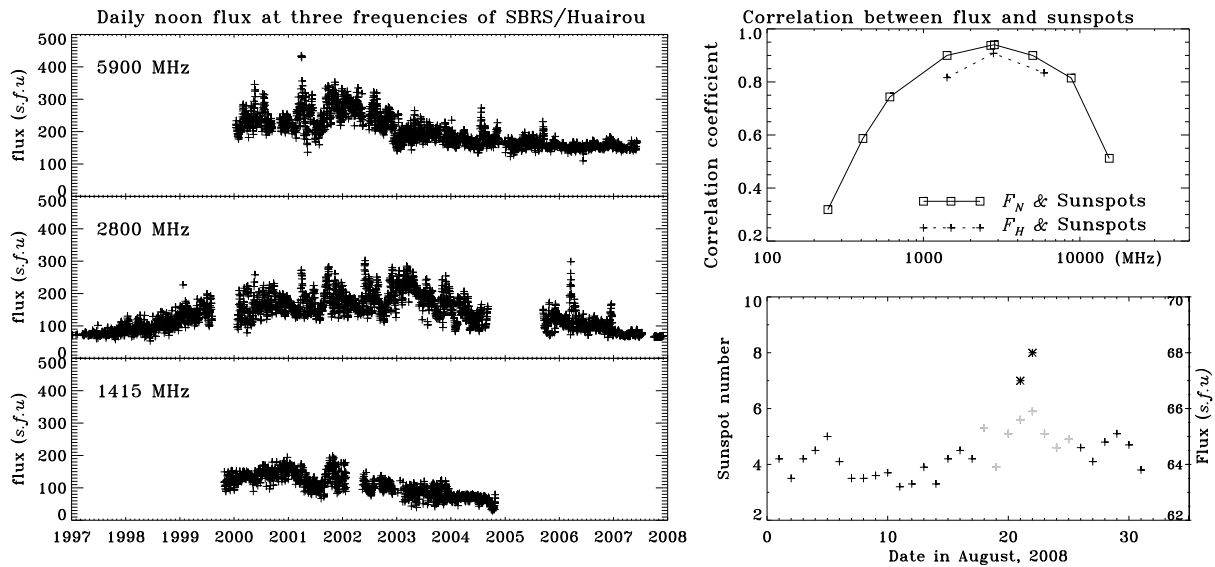} \caption{Left: daily noon flux
at 1415, 2800, and 5900 MHz observed by SBRS in solar cycle 23.
The data gap is without daily calibration or during maintenance.
Top right panel: correlation coefficients between the sunspot
numbers and the daily noon flux at various frequencies of NGDC and
SBRS. Bottom right panel: sunspots numbers (star) and the daily
noon flux (cross) at 2800 MHz in 2008 August as an example. The
gray cross is excluded because there are sunspots $\pm3$ days.
\label{figsunspots}}
\end{figure*}

The daily noon flux observed by SBRS is calibrated by equation
(\ref{calisfu}) with the calibration coefficient $C(\nu)$. At the
2.6 - 3.8 GHz band, $C(\nu)$ is $C_{TC}$. At other frequency
bands, $C(\nu)$ is the constant coefficient. The left panel of
Fig.\ref{figsunspots} plots the daily noon flux at 1415, 2800 and
5900 MHz observed by SBRS during solar cycle 23. In each band of
SBRS, we plot the same (or nearby) frequency as nine fixed
frequencies of NGDC. The top right panel plots the correlation
coefficients between the sunspot numbers and the daily noon flux
at various frequencies. The correlation coefficients are
consistent with previous works \citep{Chris51}. The bottom right
panel plots the sunspot numbers and the daily noon flux at 2800
MHz in 2008 August, as an example. There are still small sunspots
or weak active regions during August 21-22 even in the solar
minimum. In order to obtain the pure radio flux of the quiet-Sun,
we exclude the daily noon flux with sunspots $\pm3$ days (gray
cross in bottom right panel of Fig.\ref{figsunspots}). During the
solar minimum period of 2006-2009, 317 days are selected as the
candidates. The Gaussian fitness of the daily noon flux of these
317 days is decided as radio flux of the quiet-Sun statistically.
They are 55, 67, and 68 sfu at 1415, 2695, and 2800 MHz of SBRS,
respectively. The corresponding radio flux of the quiet-Sun from
NGDC are 11, 27, 35, 55, 67, 69, 118, 220, and 519 sfu at the nine
fixed frequencies, respectively.

\section{Theoretical Analysis of the Quiet-Sun Radio Emission }

The quiet-Sun radio emission is thermal radiation originated from
the ambient plasma in absence of solar activity. The mechanism is
bremsstrahlung from electrons interacting with ions in the
presence of a relatively weak magnetic field \citep{Shibasaki11}.
In an irregular propagation medium, a wave cannot be represented
by a single ray. Small fluctuations in density or magnetic field
will distort the incident plane wave as the wave phase propagates
at different speeds. \citet{Benz93} pointed out that the evidence
of scattering of solar radio emission is ambiguous, while the
scattering hypothesis has been successfully verified for pulsars
and irregularities of the interstellar medium. Moreover, other
papers (\citet{Aubier71}; \citet{Thejappa08}) concluded that (1)
the scattering effect decreases the intensity of the radio
emission and enlarges the size of the radio source, and (2) the
scattering effect increases with the radio wavelength
($\propto\lambda^{4}$) and can be neglected at shorter meter
wavelengths. We mainly studied the quiet-Sun radio emission at the
frequency range of 245 - 15400 MHz in this work; therefore, the
scattering effect can be ignored. In numerical analysis of the
quiet-Sun radio emission, most previous works are based on the
theoretical treatments (\citet{Smerd50}; \citet{Dulk85};
\citet{Benz93}). The quiet-Sun radio emission is calculated by
using equations and treatments as follows.

The radiation transfer equation is

 \begin{equation}
\frac{d(I/\mu^{2})}{ds}=\frac{\eta}{\mu^{2}}-\kappa\frac{I}{\mu^{2}}
\label{transfer}
\end{equation}

Here $I$ is the specific intensity of the radiation at frequency
of $\nu$; $\eta$ is the volume emissivity, $\kappa$ is the
absorption coefficient, and $\mu$ is the refractive index of the
medium. The optical depth and the path element have the
relationship $d\tau={\kappa}ds$. In this work, the electron
temperature $T$ can be treated as uniform in small segments
because of the entirely numerical integration. Thus, under
conditions of thermodynamic equilibrium in small segment, we have
$\eta={\mu^{2}}{\kappa}B(T)$. In the radio frequency band, the
Rayleigh-Jeans approximation is
$B(T)=\frac{2k_{b}}{c^{2}}{\nu^{2}}T$. Hence, the solution of
equation (\ref{transfer}) can be obtained:

 \begin{equation}
I=\frac{2k_{b}}{c^{2}}\mu^{2}
{\nu}^{2}T(1-e^{\tau_{0}-\tau})+I_{0}(\mu/\mu_{0})^{2}e^{\tau_{0}-\tau}
\label{Intensity}
\end{equation}

Here the intensity $I_{0}$ and refractive index $\mu_{0}$ are at
an optical depth of $\tau_{0}$. $k_{b}$ is the Boltzmann's
constant.

\begin{figure*}[ht]
\epsscale{1.0} \plotone{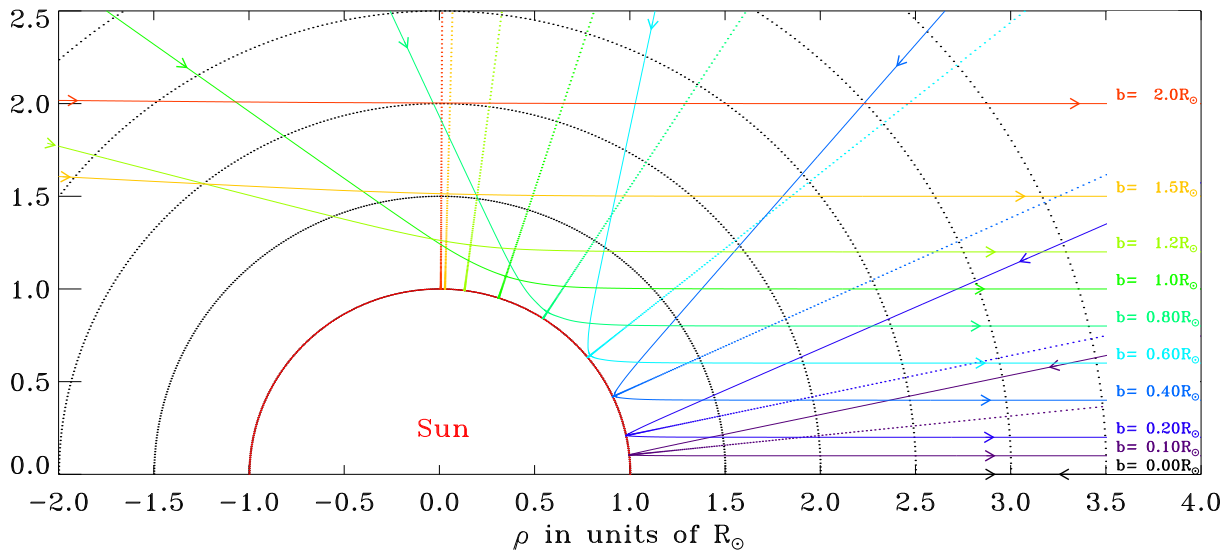} \caption{Ray trajectory of 245
MHz in polar coordinates. The bending curves of various colors
indicate the ray trajectory of various distances $b$ from the
center Sun-Earth line. The normal line of the refraction at the
turning point is plotted as a dashed line. The arrow indicate the
propagation direction of the ray to the observer. \label{figrey}}
\end{figure*}

The ray treatment of radiation \citep{Jaeger50} is based on the
refraction of the ray path in the solar atmosphere. The equation
of the path and the absorption of the rays can be deduced from
Snell's law. Fig.\ref{figrey} shows the ray trajectory in polar
coordinates ($\rho, \theta$). Here $\rho=R/R_{\odot}$ is the
distance from one point of the ray path to the solar center. These
rays will emerge from the solar atmosphere parallel to the
observer at a distance $b$ from the center Sun-Earth line. Both
$\rho$ and $b$ are in units of the sun's optical radius
$R_{\odot}$ ($6.95\times10^{10} cm$). The refractive index $\mu$
of an ionized medium decreases with increasing electron density.
It follows that a ray passing through the solar atmosphere
experiences continuous bending by refraction. The point where the
direction of propagation changes from that of decreasing $\mu$ to
that of increasing $\mu$ is referred to as the 'reflection point'
or, better, as the turning point. All these rays are calculated
with the equation (\ref{ftheta}) \citep{Jaeger50} and the solar
atmosphere model in Fontenla et al. (1993, 2011) paper. The path
element $ds$ of a trajectory $b$ is given by equation
(\ref{fpath})

\begin{equation}
\theta=\int^{\infty}_{\rho}\frac{bd{\rho}}{{\rho}\sqrt{{\mu}^{2}{\rho}^{2}-b^{2}}}
\label{ftheta}
\end{equation}

\begin{equation}
ds=R_{\odot}{\sqrt{(d{\rho})^{2}+({\rho}d{\theta})^{2}}}=\frac{R_{\odot}d{\rho}}{\sqrt{1-b^{2}/{{\mu}^{2}{\rho}^{2}}}}
\label{fpath}
\end{equation}

The optical depth can be deduced from the equation of the path
(\ref{fpath}) and the absorption of the rays. The optical depth
between two points ($\rho_{1}$ and $\rho_{2}$) is as follows:

\begin{equation}
\tau_{1,2}(b)=\int^{\rho_{2}}_{\rho_1}\frac{{\kappa}R_{\odot}}{\sqrt{1-b^{2}/{\mu^{2}}
\rho^{2}}}d\rho \label{tao}
\end{equation}

The refractive index is
$\mu^{2}\approx{1-(\frac{\omega_{p}}{\omega})^{2}\cdot(1\pm\frac{\Omega_{e}}{\omega}|cos\theta|)^{-1}}\simeq{1-(\frac{\omega_{p}}{\omega})^{2}}$
because the ray frequency is much greater than the local
gyrofrequency $\omega\gg\Omega_{e}$ along its propagation in the
solar atmosphere of the quiet-Sun \citep{Benz93}. The formulae of
absorption coefficient $\kappa$ differ in different papers. Some
papers (\citet{Smerd50}; \citet{Bracewell56}; \citet{Thejappa10})
used $\kappa$ deduced by classical collision theory. The classical
collision theory defined the absorption coefficient
$\kappa=\frac{\nu_{e,i}}{{\mu}c}(\frac{\omega_{p}}{\omega})^{2}$.
Here $\nu_{e,i}$ is the collision frequency of thermal
electrons/ions. Some papers (\citet{Dulk85}; \citet{Gary90}) used
$\kappa$ with the Gaunt factor deduced by quantum theory. The
formula of $\kappa$ in \citet{Dulk85} is the same as the
approximate analytic formula of \citet{Novikov73};
\citet{Rybicki86} at radio wavelength ($\hbar\nu<k_{b}T$). We
compare the Gaunt factor value of \citet{Dulk85} with that of
\citet{vanHoof14} at the solar atmosphere from 100 Mhz to 30 GHz
and find that the difference is of $<2\%$ which has a very small
impact on the numerical result. Thus in this work we still use the
value calculated by the equation of \citet{Dulk85}. Table
\ref{tbl-1} lists the equations and parameters from different
papers. They are all approximatively in the range of
$0.15\sim0.25\frac{N^{2}}{\nu^{2}T^{3/2}\mu}$ under the condition
of the solar atmosphere since the classical collision theory is
valid and close to the quantum theory when $\hbar\nu\ll k_{b}T$.

\begin{table*}[ht]
\caption{Parameters of collision frequency $\nu_{e,i}$, average
Gaunt Factor $\overline{g}_{ff}(\nu,T)$, and absorption
coefficient $\kappa$. \label{tbl-1}}
\begin{tabular}{crrrr}
\tableline\tableline
     1           &        2     &    3      &     4  \\
  Paper          & $\nu_{e,i}$ (cgs)  & $\overline{g}_{ff}(\nu,T)$    & $\kappa$ (cgs)\\
\tableline
 Smerd (1950)    & $\frac{1.36N}{T^{3/2}}ln[1+(\frac{4k_{b}T}{e^{2}N^{1/3}})^{2}]$ &- & $\frac{3.65\times10^{-3}N^{2}}{{\nu}^{2}T^{3/2}\mu}{ln[1+(\frac{4k_{b}T}{e^{2}N^{1/3}})^{2}]}$ \scriptsize\\
 Bracewell (1956) & $\frac{1.36N}{T^{3/2}}ln[1+(\frac{4k_{b}T}{e^{2}N^{1/3}})^{2}]$ &- & $\sim\frac{0.2N^{2}}{{\nu}^{2}T^{3/2}\mu}$ \scriptsize \\
 Dulk (1985)     & - &$\frac{\sqrt{3}}{\pi}[18.2+ln(\frac{T^{3/2}}{\nu})]$ & $\frac{9.78\times10^{-3}N^{2}}{{\nu}^{2}T^{3/2}\mu}[18.2+ln(\frac{T^{3/2}}{\nu})]$ \scriptsize $T<2\times10^{5}$     \\
 Dulk (1985)     & - &$\frac{\sqrt{3}}{\pi}[24.5+ln(\frac{T}{\nu})]$         & $\frac{9.78\times10^{-3}N^{2}}{{\nu}^{2}T^{3/2}\mu}[24.5+ln(\frac{T}{\nu})]$ \scriptsize $T>2\times10^{5}$ \\
 Thejappa (2010) & $\frac{4.36N}{T^{3/2}}[10.81+ln(\frac{T^{3/2}}{\nu})]$ &- & $\frac{2.34\times10^{-2}N^{2}}{{\nu}^{2}T^{3/2}\mu}[10.81+ln(\frac{T^{3/2}}{\nu})]$ \scriptsize
 (corona)\\
\tableline
\end{tabular}
\end{table*}

From equation (\ref{tao}) and $\kappa$ in Table \ref{tbl-1}, we
have

\begin{equation}
\tau_{1,2}(b)=\frac{9.78\times10^{-3}R_{\odot}}{\nu^{2}}{\frac{\pi}{\sqrt{3}}}\int^{\rho_{2}}_{\rho_1}\frac{{\overline{g}_{ff}(\nu,T)}{N}^{2}}{{T}^{3/2}\sqrt{\mu^{2}-b^{2}/
\rho^{2}}}d\rho
 \label{tau12}
\end{equation}

$N$ and $T$ are the electron density and electron temperature in
the solar atmosphere, respectively. They will be discussed in the
next subsection. Equation (\ref{tau12}) has a singularity point
when $\mu=b/ \rho$, $\rho = \rho_{0}$. This point was named the
turning point or 'reflection point' (\citet{Smerd50};
\citet{Bracewell56}) where the direction of propagation changes
from that of decreasing $\mu$ to that of increasing $\mu$. The
appendix proves that the integration is convergent near the
singularity point so long as the electron density $N$ increased
limitedly and monotonously along the height $\rho$. Therefore,
when $\rho > \rho_{0}$, the integration can be calculated with
numerical integration.

\subsection{The Electron Density and Temperature}

\begin{figure*}[ht]
\epsscale{1.00} \plotone{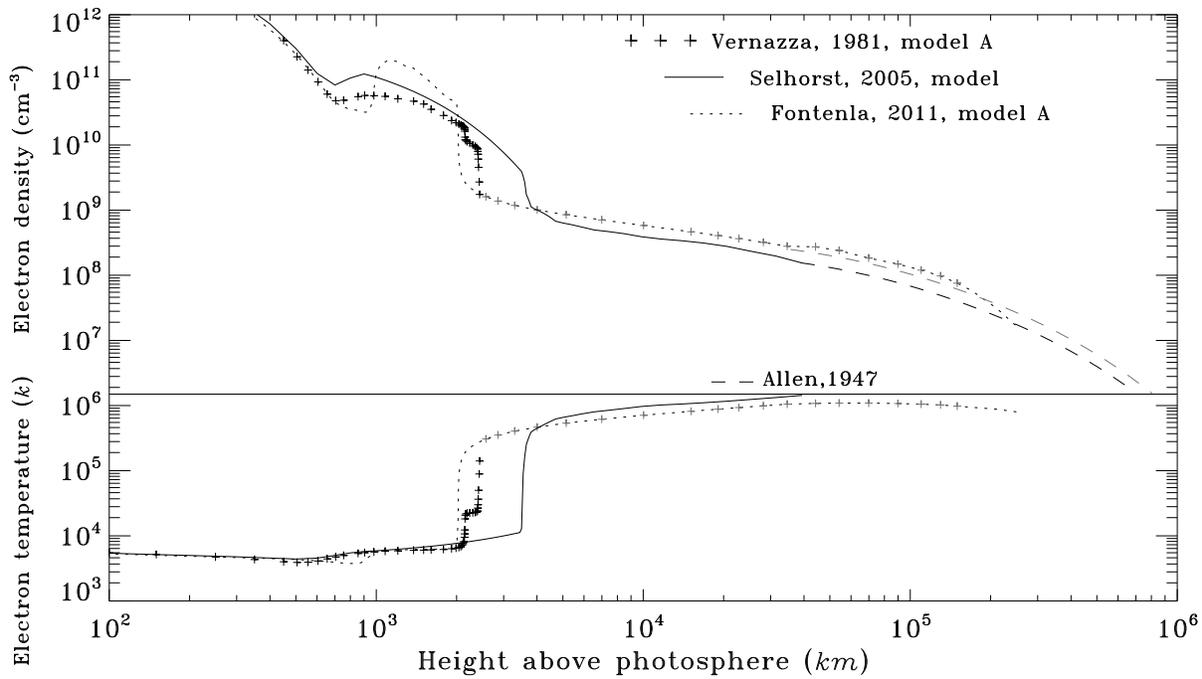} \caption{The electron density
(top panel) and electron temperature (bottom panel) distribution
of different models along the height above photosphere. The
classical model \citet{Alllen47} is plotted as long dashed line
only at height of higher than $10^{4}km$. \label{densete}}
\end{figure*}

There are many models of the electron density and electron
temperature of the quiet-Sun atmosphere. The VAL III model
\citep{Vernazza81} determined semi-empirical models for six
components of the quiet solar chromosphere in using of EUV
observations. The series of FAL models (Fontenla et al. 1993,
2009, and 2011) built the semi-empirical models with the optical
continuum and EUV/FUV observation. These two models are excellent
for reconstructing the optical and ultraviolet observations but
still have some discrepancies when describing the radio
observations. The discrepancies will be illustrated in the next
subsection. \citet{Selhorst05} proposed a hybrid model that uses
the FAL C model \citep{Fontenla93} from the photosphere to $1800
km$, \citet{Zirin91} model from $1800$ to $3500 km$ in the
chromosphere, and \citet{Gabriel92} model from $3500$ to $40000
km$ in the transition region and corona. These three models are
selected in this work because the numerical results are not far
apart from the observations. Fig.\ref{densete} shows the electron
density and electron temperature distribution of different models
along the height above the photosphere. The \citet{Fontenla11}
model is plotted as dashed line. It gave parameters from the
chromosphere to the high corona until the height of
$\sim2\times10^{5}km$. In the higher corona, the electron density
is given by the \citet{Alllen47} model as follows:

\begin{equation}
N=10^{8}(1.55\rho^{-6}+2.99\rho^{-16}).cm^{-3}
 \label{dense2}
\end{equation}

The Allen model (long-dashed line in Fig.\ref{densete}) will match
the value of the Fontenla model at the height of
$\sim2\times10^{5} km$. The Vernazza model is plotted as a cross
in Fig.\ref{densete}. It gave parameters in the chromosphere and
transit region until the height of $2439 km$. We decide to use the
same parameters as the Fontenla model in the corona and the Allen
model in the higher corona as before. They are plotted as gray
crosses in the figure. The Selhorst model is plotted as a solid
line in Fig.\ref{densete}. It gives parameters from the
chromosphere to the corona until the height of $3.93\times10^{4}
km$. We decide to use the Allen model after modification (gray
long-dashed line in Fig.\ref{densete}) in the higher corona. Then
three hybrid models that we used in this work are the FAL+Allen
(F+A) model, VAL+FAL+Allen (V+F+A) model, and Selhorst+Allen (S+A)
model. For all the models, the electron temperature in the higher
corona is equal to the last value given by the corresponding
model. We will compare the numerical results of three hybrid
models with the observations in the subsection 3.3.

\subsection{The Optical Depth and Brightness Temperature across the
Solar Disk}

\begin{figure*}[ht]
\epsscale{1.00} \plotone{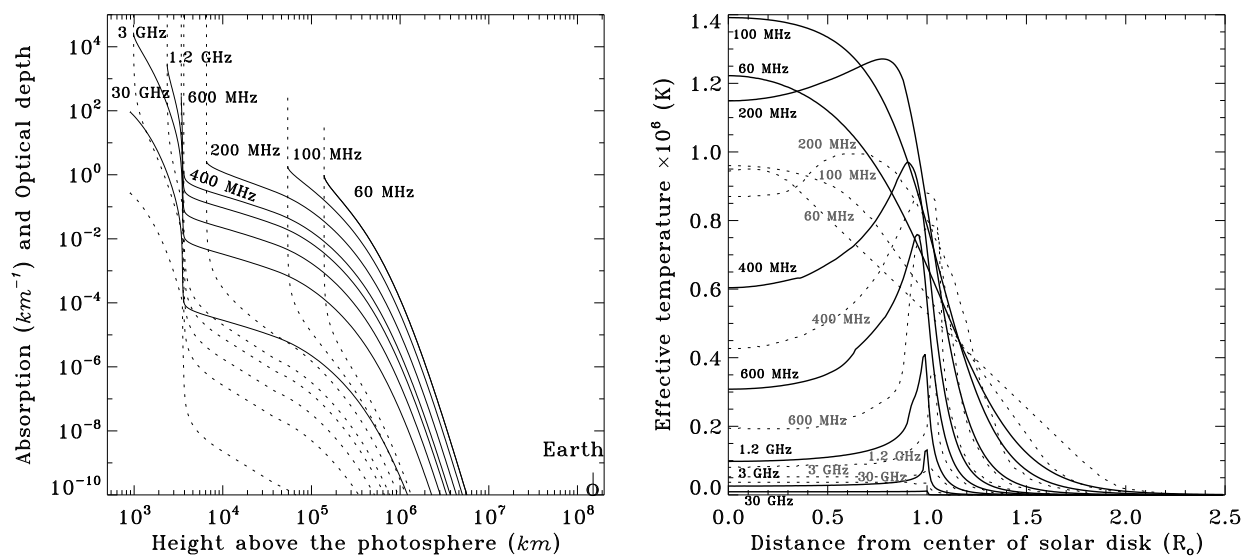} \caption{Left panel plots the
absorption $\kappa_{n}$ (long dashed line) and the optical depth
$\tau_{n,\infty}$ (real line) along the solar height, for each of
the central line at eight fixed frequencies. Right panel is the
brightness temperature across the solar disk at eight fixed
frequencies. Black lines are the numerical result with
\citet{Selhorst05}+ \citet{Alllen47} model. Dashed lines are the
results of approximation solution in
\citet{Smerd50}.\label{figbt}}
\end{figure*}

We do numerical integration entirely in this paper. In equation
(\ref{tau12}), the optical depth $\tau_{n,n+1}(b)$ is calculated
for about 500 points from the turning point (inner limit) to the
point after which the contribution is very small (outer limit).
The appendix proves that the integration of equation (\ref{tau12})
near the turning point is convergent. Beyond the outer limit, the
absorption is very small ($< 10^{-10}$ $km^{-1}$ in this work)
thus, the optical depth can be ignored. The 500 points are decided
as follows: (1) the zero point is the turning point $\rho_{0}$;
(2) assume the first point $\rho_{1}=\rho_{0}+\Delta\rho_{1}$,
$\Delta\rho_{1}\leq1R_{\odot}$, and calculate the optical depth
$\tau_{0,1}$ between the two points; (3) set the midpoint between
point 0 and point 1, $\rho_{m}=\rho_{0}+(\rho_{1}-\rho_{0})/2$,
and calculate the optical depth of $\tau_{0,m}$ and $\tau_{m,1}$,
respectively; (4) if
$|\frac{\tau_{0,m}+\tau_{m,1}-\tau_{0,1}}{\tau_{0,1}}|>10^{-2}$,
 change the midpoint to point 1 and set a new midpoint. Do this circulatory
calculation until
$|\frac{\tau_{0,m}+\tau_{m,1}-\tau_{0,1}}{\tau_{0,1}}|\leq
10^{-2}$. The calculation error is very small. Then the first
point $\rho_{1}=\rho_{m}$ is decided. Usually, the first point is
decided as $\Delta\rho_{1}=10^{-12}\sim10^{-10}R_{\odot}$ by
experience. (5) Decide the rest points with this method. It should
be taken care that the electron density and electron temperature
vary abruptly in the transition region. The step should be
$10^{-12}R_{\odot}\leq\Delta\rho_{n}\leq1.5\times\Delta\rho_{n-1}$
to avoid the big error during calculation. For the points beyond
the outer limit, the optical depths are approximated as zero. The
left panel of Fig.\ref{figbt} plots the absorption $\kappa_{n}$
and optical depth $\tau_{n,\infty}$ of the central line calculated
with the Selhorst + Allen model at eight fixed frequencies as an
example. The right panel of Fig.\ref{figbt} shows the brightness
temperature (or radiation intensity), which is calculated with
equation (\ref{Intensity}) and (\ref{tau12}) for different $b$ at
eight fixed frequencies. Black lines are the numerical result of
the Selhorst + Allen model in this work. Dashed lines are the
result of \citet{Smerd50} with the approximation solution. It
shows differences between two results.

Fig.\ref{figbt} illustrates the main generating region of the
quiet-Sun radio emission at varied frequencies. The result is
consistent in the main but a little different from the classical
results (\citet{Smerd50}; \citet{DelaLuz10}). At high frequency
($3\sim30 GHz$), the absorption ($<10^{-5} km^{-1}$) and optical
depth ($<0.01$) is very small in the corona. The emissivity
$\eta={\mu^{2}}{\kappa}B(T)$ are very small in the corona. Thus,
the thermal bremsstrahlung emission of the quiet-Sun at high
frequency ($3-30 GHz$) is mainly produced in the chromosphere. At
low frequency ($<300 MHz$), the radio emission cannot propagate
into the chromosphere. The thermal bremsstrahlung emission of the
quiet-Sun at low frequency ($<300 MHz$) is mainly produced in the
corona. The thermal bremsstrahlung emission of the quiet-Sun at
intermediate frequency ($300 MHz<\nu<3 GHz$) is produced in both
the chromosphere and the corona. The limb brightening appears
obviously at a frequency range of $200 MHz<\nu\leq30 GHz$. The
brightness temperature of the solar center radio emission is
consistent with the electron temperature of the main generating
region.

\subsection{Brightness Temperature and Flux Density of the Quiet-Sun}

\begin{figure*}[ht]
\epsscale{1.00} \plotone{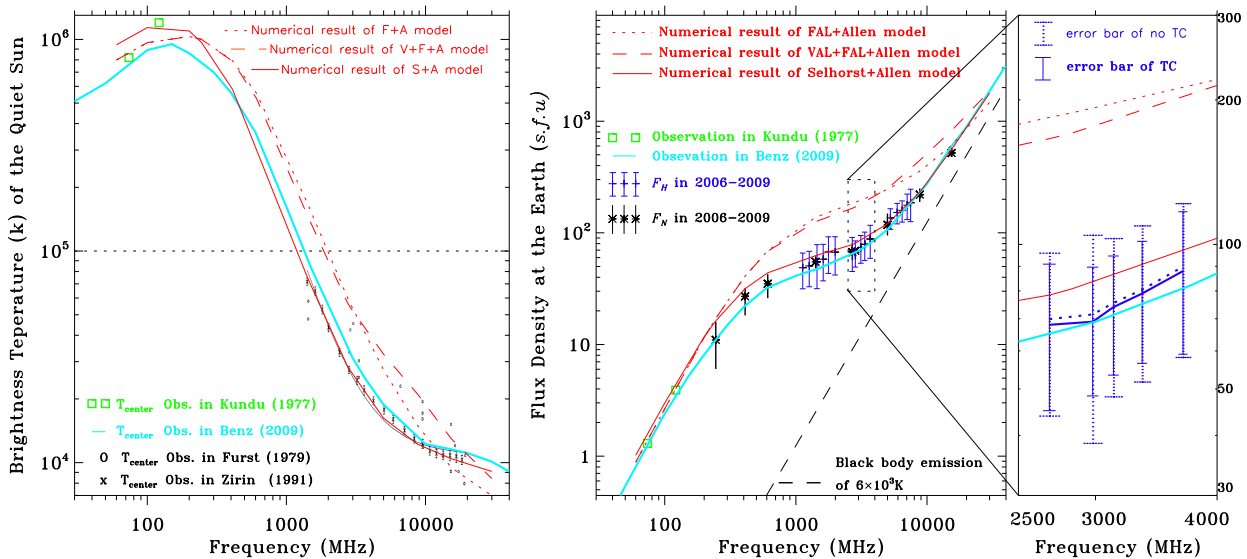}
 \caption{Left: brightness temperature spectrum of the center line
of the quiet-Sun. The errors bar are derived from the observations
in \citet{Zirin91}. Middle: comparison between numerical results
and observations of flux spectrum of the quiet-Sun. The error bars
are derived from observations of SBRS and NGDC, respectively. The
observation result in \citet{Benz09} has an accuracy of about
10\%. Right: enlargement of a small rectangle from the middle
panel. It shows the observation fluxs (thick blue lines) and error
bars (thin blue lines) of TC and no TC in the range of 2500-4000
MHz as an example.\label{figobsbte}}
\end{figure*}

We calculate the brightness temperature spectrum and flux spectrum
of the quiet-Sun radio emission for three hybrid models (F+A,
V+F+A, S+A) at various frequencies. The total amount of radiation
per unit time, unit frequency interval, and unit angle from the
Sun to a distant observer is given by

\begin{equation}
E=2\pi R_{0}^{2} \int_{0}^{\infty}I(d)d\delta d\label{flux},
\end{equation}

in units of $erg.cm^{-2}Hz^{-1}s^{-1}$. In practice, the upper
limit of this integral is replaced by a finite value $d_{p}$, out
of which the radiation contribution is very small. The flux
density of the quiet-Sun can be transformed with the conversion $1
s.f.u=10^{-22}Wm^{2}Hz^{-1}=10^{-19}erg.cm^{-2}Hz^{-1}s^{-1}$.
Then we compare the numerical results with observations and find
the difference. The left panel of Fig.\ref{figobsbte} shows the
comparison of brightness temperature spectrum between the
observations and numerical results of different models. We chosse
the observations of \citet{Benz09} and \citet{Zirin91} as the
standard spectrum because the flux spectrum of \citet{Benz09} is
well consistent with the observations of SBRS-/HSO and NGDC
(middle panel of Fig.\ref{figobsbte}). At frequency of $15400MHz$
or nearby, the numerical results of all models and all the
observations (\citet{Fuerst80}; \citet{Zirin91}; \citet{Benz09})
are consistent. This indicates that the parameters in the low
chromosphere of all the models fit the observation well. At
frequency of $3-10GHz$, the numerical results of the F+A model
(mainly Fontenla model) and V+F+A model (mainly Vernazza model)
are close to the observations of \citet{Fuerst80}, while the
numerical results of the Selhorst+Allen model are close to the
observations of \citet{Zirin91} and \citet{Benz09}. The radio
emission at the frequency of $300MHz-3GHz$ comes from the
transition region and corona. The numerical results of the
Selhorst+Allen model (red solid line) are close to the
observations of Benz (cyan solid line) well. Thus, we think that
from the transition region to the low corona the parameters of the
Selhorst model fit the observations well. It is complex to
estimate wether the parameters in the higher corona are good or
not owing to a lack of observations at low frequency ($<300MHz$).

\begin{table*}[ht]
\caption{The center-line brightness temperature and flux density
of the quiet-Sun at various frequencies}
\begin{tabular}{crrrrrrrrrrr}
\tableline\tableline
$\nu[MHz]$&$T_{c}[K]$&$T_{c}[K]$&$T_{c}[K]$&$T_{c}[K]$&$F_{\odot}$&$F_{\odot}$&$F_{\odot}$&$F_{\odot}$&$F_{\odot}$&$F_{\odot}$\\
          &   $F+A$  & $V+F+A$&  $S+A$   &$Benz$    &$F+A$      & $V+F+A$ & $S+A$     & $Benz$    &  $F_{N}$  &  $F_{H}$  \\
\tableline
 245   &$1.00\times10^{6}$&$1.00\times10^{6}$&$9.69\times10^{5}$&$7.88\times10^{5}$&  17.3     &  17.3     &  15.7     &  11.2     &  11  &    \scriptsize \\
 410   &$7.93\times10^{5}$&$7.89\times10^{5}$&$5.90\times10^{5}$&$5.50\times10^{5}$&  41.4     &  21.3     &  30.3     &  22.2     &  27  &    \scriptsize \\
 610   &$5.56\times10^{5}$&$5.27\times10^{5}$&$2.84\times10^{5}$&$3.58\times10^{5}$&  69.8     &  68.9     &  41.9     &  32.3     &  35  &    \scriptsize \\
 1415  &$1.68\times10^{5}$&$1.48\times10^{5}$&$7.43\times10^{4}$&$9.70\times10^{4}$&  139      &  127      &  60.8     &  46.9     &  55  &  55 \scriptsize \\
 2695  &$6.21\times10^{4}$&$5.85\times10^{4}$&$2.93\times10^{4}$&$4.19\times10^{4}$&  185      &  167      &  77.9     &  64.7     &  67  &  67 \scriptsize \\
 2800  &$5.76\times10^{4}$&$5.49\times10^{4}$&$2.76\times10^{4}$&$3.68\times10^{4}$&  189      &  172      &  80.3     &  66.8     &  69  &  68 \scriptsize \\
 4995  &$2.74\times10^{4}$&$3.25\times10^{4}$&$1.62\times10^{4}$&$1.88\times10^{4}$&  246      &  251      &  121      &  107      &  118 &    \scriptsize \\
 8800  &$1.49\times10^{4}$&$2.10\times10^{4}$&$1.22\times10^{4}$&$1.38\times10^{4}$&  359      &  450      &  233      &  235      &  220 &    \scriptsize \\
 15400 &$9.46\times10^{3}$&$1.29\times10^{4}$&$1.04\times10^{4}$&$1.16\times10^{4}$&  612      &  832      &  562      &  599      &  519 &    \scriptsize \\
\tableline
\end{tabular}
\tablecomments{The first column is frequency. The next three
columns are numerical results of center-line brightness
temperature with different models (FAL+Allen; VAL+FAL+Allen;
Selhorst+Allen). The fifth column is center-line brightness
temperature in \citet{Benz09}. From sixth to eighth columns are
numerical results of flux density with different models. The last
three column are observation results of flux density of the
quiet-Sun (\citet{Benz09}; NGDC online data; and
SBRS).\label{tbl-2}}
\end{table*}

The middle panel of Fig.\ref{figobsbte} plots the flux spectrum of
the quiet-Sun radio emission. The quiet-Sun fluxes of SBRS
observations (Gaussian fitness value) and the error bars
($3\sigma$) are plotted as blue crosses and short vertical lines.
The quiet-Sun fluxes of NGDC (Gaussian fitness value) are plotted
as black stars. The NGDC did not present the error of solar flux.
Here we use three times the standard deviation ($3\sigma$) of the
quiet-Sun data from NGDC as error bars (black short vertical
lines). The standard deviation is calculated separately at the low
value part and high value part. We find that both the quiet-Sun
flux spectra of SBRS and NGDC fit the observation of Benz (cyan
solid lines) well. Thus, our work agrees with the brightness
temperature spectrum and flux spectrum of \citet{Benz09} as the
standard spectrum of the quiet-Sun radio emission. All the values
of brightness temperature and flux density of the quiet-Sun at
various frequencies are listed in table \ref{tbl-2}. The right
panel of Fig.\ref{figobsbte} shows the observation flux and error
bars of TC (solid blue line) and no TC (dashed blue line) in the
range of 2500 - 4000 MHz as an example. It indicates that
calibration of TC is more accurate than that of no TC. The red
lines are the same numerical results of solar models as in the
middle panel of Fig.\ref{figobsbte}. The cyan line is the
observation of Benz. As analyzed in subsection 2.3, the difference
between TC and TWC calibration is very small ($\sim2\%$). The
observation flux and error bars of TWC are not plotted together.

\section{Conclusion and Discussion}

SBRS has been observing the Sun and obtaining plentiful data since
1994. This work adopted the observations of SBRS to investigate
the calibration procedure and study the quiet-Sun radio emission.
The calibration gives accurate observations which is basic and
important in the study of the quiet-Sun radio emission, while the
study of the quiet-Sun radio emission is a scientific extension of
the former part. We first study various impacts to the calibration
results and conclude that:

1) Generally, the calibration coefficient is constant and should
be upgraded annually. Actually, the calibration result with
constant coefficient is found to be related with the local air
temperature at all frequency bands of SBRS. One possible reason is
that the electronic apparatus of the instruments are influenced by
the local air temperature. Thus, the correlation between
calibration and local air temperature will vary if there is an
adjustment to the instrument.

2) The relationship between calibration results (panel (c) of
Fig.\ref{figcali1}) and the humidity (panel (a) of
Fig.\ref{figcali1}) or other weather conditions is not clear.
There is no distinct evidence that the calibration was influenced
by normal weather conditions except for air temperature.

3) The Sun elevation has a small effect on the calibration of this
work. The absorption of the atmosphere should be considered only
when the frequency is higher than 17 GHz \citep{Tsuchiya65} or the
Sun elevation is lower than $10^{o}$.

The calibration accuracy is also influenced by the occasional
abnormal signal. Some possible reasons for the abnormal signal are
listed as follows:

1) Various kinds of interference: wireless communication, plane
and airport, satellite, vehicle, lightning, etc. These will result
in large values.

2) Unstability of the feed, cable, or noise source.

3) Sometimes, the gain factor may be out of normal range when
there is a strong signal \citep{Yan02}. The beam of the antenna
will offset the sun center if the tracking control is out of
normal range. These will result in big or saturated values, or
small values. The relationship between the abnormal signal and bad
weather conditions is still not clear. The calibration with
equation (\ref{calisfu}) will eliminate the frequency property of
gain factor $G_{r}(\nu)$, and flatten the frequency property of
the antenna system with $C(\nu)$ of each frequency. The
calibration with equation (\ref{calisfu}) does not need the values
of $A_{e}(\nu)$ and ${T_{n}(\nu)-T_{t}(\nu)}$.

In order to improve the calibration accuracy, we investigate the
influence and properties of the instrument from the data analysis
and comparison between different calibration coefficients. All the
investigations are under the fundamental of calibration. First, we
exclude the abnormal data that are not well observed or are
influenced by the interference. Then we compare the calibration
results of four sets of calibration coefficients, including
average value ($\overline{C}$), the Gaussian fitness value
($\widehat{C}$), constant coefficients after temperature
correction ($C_{TC}$), and constant coefficients after
temperature-wavelet correction ($C_{TWC}$). The main analysis
results are as follows: (1) In the 2.6 - 3.8 GHz band, the
calibration errors $\sigma_{G}$ are smaller than $\sigma_{av}$ at
$82\%$ of frequencies. Comparing with other constant coefficients
of calibration, $\widehat{C}$ is the best. (2) At 2800 MHz,
$\sigma_{TC}$ or $\sigma_{TWC}$ are about $10\%-20\%$ smaller than
$\sigma_{G}$ at various observation terms. The RSDs of
$\sigma_{TC}$ and $\sigma_{TWC}$ are less than $10\%$, while RSDs
of $\sigma_{G}$ are greater than $10\%$ sometimes because of the
influence of the temperature. The $C_{TC}$ is used in the
calibration in the years of 2008 and 2009. The calibration error
$\sigma_{TC}=6.9 s.f.u$ is about the same as that in the years of
2004 - 2007. (3) The calibrations of SBRS1 and SBRS3 also have a
strong relationship with the local air temperature. But there is
no significant difference between TC (or TWC) calibration and
constant calibration of SBRS1 and SBRS3, which were examined and
repaired many times. $\widehat{C}$ is used in the calibration for
several short observation terms. The standard deviations of
calibration are varied within $8-12 s.f.u$ at 1415 MHz and within
$5-18 s.f.u$ at 5900 MHz for different observation terms,
respectively. The observation term with the smallest standard
deviation indicates a stable system and few occasional
interferences, therefore signifying the best status of the
instrument.

The daily noon fluxes of SBRS are calibrated with improved
$C(\nu)$. We select the daily noon flux without sunspots $\pm3$
days as the flux of the quiet-Sun during the solar minimum period
of 2006 - 2009. The spectrum of daily noon flux observed by SBRS
approaches the spectrum from NGDC well. The daily noon flux is
correlated with sunspot numbers. The result is consistent with
previous works \citep{Chris51}.

Based on the above investigations, we further study the radio
emission from the quiet-Sun. The flux spectrum is strictly the
observation without sunspots during the solar minimum period of
2006 - 2009. The numerical simulation of the quiet-Sun radio
emission in this work is improved with several semi-empirical
solar models and entirely numerical integration instead of
approximate solution. The theoretical and treatment has been
testified well because it can deduce the same result as that of
many papers (\citet{Jaeger50}; \citet{Thejappa92};
\citet{Thejappa10}; etc). Different works have different
treatments, such as (1) different approximations of radiation
transformations, (2) different absorption coefficients $\kappa$,
(3) consideration of refractive index, scattering effect, or
magnetic field. This work uses the radiation transfer equation and
bremsstrahlung mechanism of emission without approximation and
without scattering effect and magnetic field. The ray treatment of
radiation is based on the refraction of the ray path in the solar
atmosphere \citep{Jaeger50}. The refractive index $\mu$ and
absorption coefficient $\kappa$ in \citet{Dulk85} are considered
since the Gaunt factor in \citet{Dulk85} is very close to the
numerical value of \citet{vanHoof14}.

The VAL \citep{Vernazza81} and FAL \citep{Fontenla11} models are
excellent for reconstructing the optical and ultraviolet
observations but still have some discrepancies in describing the
radio observation. \citet{Zhang01} discuss the discrepancy and
attribute the difference to an underestimation of Fe abundance
used in the calculation of the UV line emission. Thus,
\citet{Selhorst05} proposed a hybrid model, of which the
transition region is about 1000 $km$ higher than in typical VAL
and FAL models. We regulate solar models with a combination of
different models and compare the numerical results with the
observations. The numerical results of the models should be within
the confines (error bars) of observations. Finally, we find an
appropriate hybrid model that uses the Selhorst model from the
chromosphere to the height of $3.93\times10^{4} km$ above the
photosphere and the Allen model in the higher corona. The
comparison (Fig.\ref{figobsbte}) between the numerical results and
the observations indicates the followings: (1) The numerical
results of all models and all observations are consistent at 15400
MHz or nearby. The solar models in the low chromosphere are
unanimous. The optical depth at 15400 MHz is very small in the
transition region or higher no matter what model is selected. (2)
The argument happened at the transition region or nearby, which is
the main emission region at $300 MHz-10 GHz$. Only the flux
spectrum of the Selhorst model is within the error bars of the
observation. The theoretical results of the VAL and FAL models are
about two times larger than observations. At the frequency of $300
MHz-10 GHz$ the optical depth varied mainly within $0.01<\tau<10$
in the transition region or higher. It is difficult to modulate
the solar models only with radio observations because each
modulation in this region will impact on the results of the whole
band of $300 MHz-10 GHz$. But the radio observations will help us
to validate the models we known. (3) In the corona, the numerical
results of the Allen model are close to the observations at the
frequency of $100-300 MHz$. There is small argument on the solar
models in the corona. But in this work we can not assure exactly
the parameters in the corona as a result of few observations and
the scattering effect in low frequency ($<300 MHz$). (4) The
observation with higher accurate calibration will help us further
qualify the empirical/semi-empirical solar models. The solar model
can be decided when the theoretical result is within the error.

The brightness temperature distribution across the solar disk in
this work (right panel of Fig.\ref{figbt}) shows that the limb
brightening happens in the frequency range of $>200 MHz-30 GHz$.
\citet{Kundu77} showed that no limb brightening happens at the
frequency of lower than $\sim121 MHz$. Mercier \& Chambe (2009,
2012) studied the radio images of the quiet-Sun at $150-450 MHz$
but did not discuss limb brightening. However, their papers give
us some clue that the limb brightening increased with the
frequency in the range of $300-450 MHz$. Some works also found the
limb brightening at a high frequency of $5GHz$ \citep{Kundu79} and
near the solar polar at $17GHz$ \citep{Selhorst10}. The result of
higher frequency range ($>30 GHz$) was discussed in
\citet{DelaLuz11}. Please note that the limb brightening of the
numerical results is higher than in observations. This may be
attributed to the spicules, which will decrease the limb
brightening \citep{Elzner76}. Another reason is perhaps no
consideration of fluctuation, scattering effect, and magnetic
field. We need more work and more observations to understand this
problem. The new construction of the Chinese solar radio
heliograph \citep{Yan09} will provide more observations on the 2D
image of the Sun at a wide frequency range of $400 MHz - 15 GHz$.
In a word, this work gives an appropriate numerical method on the
study of the quiet-Sun radio emission.

\acknowledgments The authors thank the referee for helpful and
valuable comments on this paper. This work is supported by NSFC
grants 11103044, 11221063, 11273030, 11373039, 41175016, 41375041,
and 41375057; MOST grants 2011CB811401 and 2014FY120300; and the
National Major Scientific Equipment R\&D Project ZDYZ2009-3.
Thanks to NGDC for the online data.

\appendix

\section{Appendix}
Here we discuss the the convergence of the integration in equation
(\ref{tau12}). At the turning point $\rho_{0}$, $\mu_{0}=d/
\rho_{0}$, $\mu_{0}$ is the refractive index at $\rho = \rho_{0}$.
For $\rho = \rho_{0}+ d \rho $, when $d \rho \rightarrow 0$, it
can be proved that
$\frac{N^{2}d\rho}{\sqrt{\mu^{2}-b^{2}/\rho^{2}}}\rightarrow0$.
The refractive index $\mu^{2}=1-\frac{e^{2}N^{2}}{\pi m f^{2}}$
\citep{Benz93}, where $N=N_{0}+dN$ $cm.^{-3}$, $N_{0}$ is the
electron density at $\rho = \rho_{0}$. So,
$\mu^{2}=1-\frac{e^{2}(N_{0}+dN)}{\pi m f^{2}}$. Let the constant
$C_{1}=\frac{e^{2}}{\pi m f^{2}}$. Thus,
 \begin{equation}
\frac{N^{2}d\rho}{\sqrt{\mu^{2}-b^{2}/\rho^{2}}}=\frac{N^{2}d\rho}{\sqrt{1-C_{1}(N_{0}+dN)-b^{2}/\rho^{2}}}
\end{equation}
Presuming $dN>0$ when $d \rho>0$, we have $\frac{1}{\rho_{0}+ d
\rho} < \frac{1}{\rho_{0}}$ and $N_{0}+dN<N_{0}$. Thus A1 can be
transformed as
\begin{equation}
\frac{N^{2}d\rho}{\sqrt{1-C_{1}(N_{0}+dN)-b^{2}/\rho^{2}}} <
\frac{N^{2}d\rho}{\sqrt{1-C_{1}N_{0}-b^{2}/\rho^{2}}}
\end{equation}
As $1-C_{1}N_{0}-b^{2}\rho_{0}^{-2}=0$, and
$\frac{1}{\rho^{2}}=\frac{1}{(\rho_{0}+ d
\rho)^{2}}=\frac{1}{(\rho_{0})^{2}}-\frac{2\rho_{0}d \rho+(d
\rho)^{2}}{(\rho_{0})^{2}(\rho_{0}+d\rho)^{2}}$. Then
\begin{equation}
\frac{N^{2}d\rho}{\sqrt{1-C_{1}N_{0}-b^{2}/\rho^{2}}}=
\frac{N^{2}d\rho}{\sqrt{d^{2} \frac{2\rho_{0}d \rho+(d
\rho)^{2}}{(\rho_{0})^{2}(\rho_{0}+d\rho)^{2}}}}=\frac{\rho_{0}(\rho_{0}+d\rho)N^{2}\sqrt{d\rho}}{d\sqrt{2\rho_{0}+d
\rho}}
\end{equation}
When $d \rho \rightarrow 0$, $\rho_{0}+d\rho \rightarrow
\rho_{0}$, and $N \rightarrow N_{0}$. So
\begin{equation}
\frac{N^{2}d\rho}{\sqrt{\mu^{2}-b^{2}/\rho^{2}}}<\frac{\rho_{0}(\rho_{0}+d\rho)N^{2}\sqrt{d\rho}}{d\sqrt{2\rho_{0}+d\rho}}
<\frac{\rho_{0}(2\rho_{0}+d\rho)N^{2}\sqrt{d\rho}}{d\sqrt{2\rho_{0}+d\rho}}
<\frac{\rho_{0}(N_{0})^{2}\sqrt{2\rho_{0}d\rho}}{d} \rightarrow 0
\end{equation}
That is, the integration of equation (\ref{tau12}) near the
turning point (singularity point) is convergent so long as the
electron density $N$ increased limitedly and monotonously along
the height.

\clearpage


\begin{thebibliography}{}
\bibitem[Allen(1947)]{Alllen47} Allen, C. W. 1947, MNRAS, 107, 426
\bibitem[Allen(1957)]{Alllen57} Allen, C. W. 1957, IAUS, 4, 253
\bibitem[Aubier et al.(1971)]{Aubier71} Aubier, M., Leblanc, Y., \& Boischot, A. 1971, A\&A, 12, 435
%\bibitem[Basri et al.(1979)]{Basri79} Basri, G. S., Linsky, J. L., Bartoe, J.-D. F., et al. 1979, ApJ, 230, 924
\bibitem[Bastian et al.(1996)]{Bastian96} Bastian, T. S., Dulk, G. A., \& Leblanc, Y. 1996, ApJ, 473, 539
\bibitem[Benz et al.(1991)]{Benz91} Benz, A.O., Gudel, M., Isliker, H., et al. 1991, SoPh, 133, 385
\bibitem[Benz(1993)]{Benz93} Benz, A. O. 1993, Plasma Astrophysics: Kinetic Processes in Solar and Stellar Coronae  (Astrophysics and Space Science Library, Vol. 184; Dordrecht: Kluwer)
\bibitem[Benz(2009)]{Benz09} Benz, A. O. 2009, LanB, 4B, 4116
\bibitem[Bracewell \& Preston (1956)]{Bracewell56} Bracewell, R. N. \& Preston, G.W. 1956, ApJ, 123, 14.
\bibitem[Broten \& Medd (1960)]{Broten60} Broten, N.W., \& Medd, W.J. 1960, ApJ, 132, 279
\bibitem[Christiansen \& Hindman (1951)]{Chris51} Christiansen, W. N., \& Hindman, J.V., 1951, Nature, 167, 4251, 635.
\bibitem[DelaLuz et al.(2010)]{DelaLuz10} Dela Luz, V., Lara, A., Mendoza-Torres, J.E., et al. 2010, ApJS, 188, 437
\bibitem[DelaLuz et al.(2011)]{DelaLuz11} Dela Luz, V., Lara, A., \& Raulin, J.-P. 2011, ApJ, 737, 1
\bibitem[Dulk(1985)]{Dulk85} Dulk, G. A. 1985, ARA\&A,23, 169
\bibitem[Elzner(1976)]{Elzner76} Elzner, L. R., 1976, A\&A, 47, 9E
\bibitem[Findlay(1966)]{Findlay66} Findlay, J. W. 1966, ARA\&A, 4, 77
\bibitem[Fontenla et al.(1993)]{Fontenla93} Fontenla, J. M., Avrett, E.H., \& Loeser, R. 1993, ApJ, 406, 319
\bibitem[Fontenla et al.(2009)]{Fontenla09} Fontenla, J. M.; Curdt, W.; Haberreiter, M.; et al. 2009, ApJ, 707, 482
\bibitem[Fontenla et al.(2011)]{Fontenla11} Fontenla, J. M., Harder, J., Livingston, W., et al. 2011, JGRD, 1162, 0108
\bibitem[Fu et al.(2004)]{Fu04} Fu, Q.J., Ji, H.R., Qin, Z.H., et al. 2004, SoPh, 222, 167
\bibitem[Fu et al.(1995)]{Fu95} Fu, Q.J., Qin, Z.H., Ji, H.R., et al. 1995, SoPh, 160, 97
\bibitem[Fuerst(1980)]{Fuerst80} Fuerst, E., 1980, IAUS, 86, 25F
\bibitem[Gabriel(1992)]{Gabriel92} Gabriel, A. 1992, in NATO ASI Ser. C, 373, The Sun: A Laboratory for Astrophysics, ed. J. T. Schmelz \& J. C. Brown (Dordrecht: Reidel), 277
\bibitem[Gary et al.(1990)]{Gary90} Gary, D. E., Zirin, H., \& Wang, H.M. 1990, ApJ, 355, 321
\bibitem[Jaeger \& Westfold (1950)]{Jaeger50} Jaeger, J. C., \& Westfold, K. C., 1950, AuSRA, 3..376
\bibitem[Ji et al.(2005)]{Ji05} Ji H.R., Fu Q.J., Yan Y.H., et al. 2005, ChJAA, 5, 433
\bibitem[Jiricka et al.(1993)]{Jiricka93} Jiricka, K., Karlicky, M., Kepka, O., \& Tlamicha, A. 1993, SoPh, 147, 203
\bibitem[Kundu(1965)]{Kundu65} Kundu, M. R. 1965, Solar Radio Astronomy (New York: Interscience)
\bibitem[Kundu et al.(1977)]{Kundu77} Kundu, M. R., Erickson, W. C., \&, Gergely, T. E. 1977, SoPh, 53, 489
\bibitem[Kundu et al.(1979)]{Kundu79} Kundu, M. R., Rao, A. P., Erskine, F. T., \& Bregman, J.D. 1979, ApJ, 234, 1122
\bibitem[Martyn(1946)]{Martyn46} Martyn, D.F. 1946, Nature, 158, 632
\bibitem[Mercier \& Chambe (2009)]{Mercier09} Mercier, C., \& Chambe, G. 2009, ApJ, 700, 137
\bibitem[Mercier \& Chambe (2012)]{Mercier12} Mercier, C., \& Chambe, G. 2012, A\&A, 540, 18
\bibitem[Messmer et al.(1999)]{Messmer99} Messmer, P., Benz, A. O., \& Monstein, C. 1999, SoPh, 187, 335
\bibitem[Nelson et al.(1985)]{Nelson85} Nelson, G. J.; Sheridan, K. V.; Suzuki, S. 1985, Solar Radiophysics: Studies of emission from the sun at metre wavelengths, ed. McLean, D. J.; Nelson, G. J.; Dulk, G. A. (Cambridge and New York: Cambridge University Press), p. 113-154.
\bibitem[Novikov \& Thorne (1973)]{Novikov73} Novikov, I. D., \& Thorne, K. S. 1973, in Black Holes (Les Astres Occlus), ed. C. Dewitt, \& B. S. Dewitt (Paris: Gordon and Breach), 343
\bibitem[Pawsey \& Yabsley (1949)]{Pawsey49} Pawsey, J. L., \& Yabsley, D. E. 1949, AuSRA, 2, 198
\bibitem[Rybicki \& Lightman (1986)]{Rybicki86} Rybicki, G. B., \& Lightman, A. P. 1986, Radiative Processes in Astrophysics (New York: Wiley)
\bibitem[Sawant et al.(2001)]{Sawant01} Sawant, H.S., Subramanian, K.R., Faria, C., et al. 2001, SoPh, 200, 167
\bibitem[Selhorst et al.(2005)]{Selhorst05} Selhorst, C. L., Silva, A. V. R., \& Costa, J. E. R. 2005, A\&A, 433, 365S
\bibitem[Selhorst et al.(2010)]{Selhorst10} Selhorst, C. L., Gim¨¦nez de Castro, C. G., Varela Saraiva, A. C., et al. 2010, A\&A, 509A, 51S
\bibitem[Shibasaki et al.(2011)]{Shibasaki11} Shibasaki, K., Alissandrakis, C. E., \& Pohjolainen, S. 2011, SoPh, 273, 309S.
\bibitem[Smerd(1950)]{Smerd50} Smerd, S. F. 1950, AuSRA, 3, 34
%\bibitem[Smerd \& Westfold (1949)]{Smerd49} Smerd, S. F., \& Westfold, K. C. 1949, Phil. Mag., 40, 831.
\bibitem[Sych \& Yan(2002)]{Sy02} Sych, R. A., \& Yan, Y. H. 2002, ChJAA, 2, 183S.
\bibitem[Tan et al.(2009)]{Tan09} Tan, C.M., Yan Y.H., Tan, B.L., \& Xu G.L. 2009, ScChG, 52, 1760
\bibitem[Tanaka et al.(1973)]{Tanaka73} Tanaka, H., Castelli J. P., Covington A. E., et al. 1973, SoPh, 29, 243-262
\bibitem[Thejappa \& Kundu(1992)]{Thejappa92} Thejappa, G., \& Kundu, M. R. 1992, SoPh, 140, 19T
\bibitem[Thejappa \& MacDowall (2008)]{Thejappa08} Thejappa, G., \& MacDowall, R. J. 2008, ApJ, 676, 1338T.
\bibitem[Thejappa \& MacDowall (2010)]{Thejappa10} Thejappa, G., \& MacDowall, R. J. 2010, ApJ, 720, 1395T
\bibitem[Tsuchiya \& MacDowall (1965)]{Tsuchiya65} Tsuchiya, A., \& Nagame, K. 1965, PASJ, 17, 86T.
\bibitem[vanHoof et al.(2014)]{vanHoof14} van Hoof, P. A. M.; Williams, R. J. R.; Volk, K.; et al. 2014, MNRAS, 444, 420V.
\bibitem[Vernazza et al.(1981)]{Vernazza81} Vernazza, J. E., Avrett, E. H., \& Loeser, R. 1981, ApJS, 45, 635V.
\bibitem[Yan et al.(2002)]{Yan02} Yan, Y.H., Tan, C.M., Xu, L., et al. 2002, ScChA, 45(Supp), 89-96
\bibitem[Yan et al.(2009)]{Yan09} Yan, Y.H., Zhang, J., Wang, W., et al. 2009, EM\&P., 104, 97Y.
\bibitem[Zhang et al.(2001)]{Zhang01}Zhang, J.; Kundu, M. R.; White, S. M.; et al. 2001, ApJ, 561, 396Z
\bibitem[Zirin et al.(1991)]{Zirin91} Zirin, H., Baumert, B. M., \& Hurford, G. J. 1991, ApJ, 370, 779Z
\end{thebibliography}
\end{document}